# Population of Exciton-Polaritons *via* Luminescent *sp³* Defects in Single-Walled Carbon Nanotubes

*Jan M. Lüttgens, Felix J. Berger, Jana Zaumseil\**

Institute for Physical Chemistry and Centre for Advanced Materials, Universität Heidelberg, D-69120 Heidelberg, Germany

ABSTRACT

Semiconducting single-walled carbon-nanotubes (SWCNTs) are an interesting material for strong-light matter coupling due to their stable excitons, narrow emission in the near infrared and high charge carrier mobilities. Furthermore, they have emerged as quantum light sources as a result of the controlled introduction of luminescent quantum defects ($sp^3$-defects) with red-shifted transitions that enable single-photon emission. The complex photophysics of SWCNTs and overall goal of polariton condensation pose the question of how exciton-polaritons are populated and how it might be optimized. The contributions of possible relaxation processes, i.e., scattering with acoustic phonons, vibrationally assisted scattering, and radiative pumping, are investigated using angle-resolved reflectivity and time-resolved photoluminescence measurements on microcavities with a wide range of detunings. We show that the predominant population mechanism for SWCNT exciton-polaritons in planar microcavities is radiative pumping. Consequently, the limitation of polariton population due to the low photoluminescence quantum yield of nanotubes, can be overcome by luminescent $sp^3$ defects. Without changing the polariton branch structure, radiative pumping through these emissive defects leads to an up to 10-fold increase of the polariton population for detunings with a large photon fraction. Thus, the controlled and tunable functionalization of SWCNTs with $sp^3$ defects presents a viable route towards bright and efficient polariton devices.

Exciton-polaritons are part-light, part-matter quasiparticles that form when an exciton interacts strongly with a cavity photon such that the energy exchange between them is faster than the decay of the separate components. They have attracted much attention for their unique properties, e.g., the ability to form non-equilibrium Bose-Einstein condensates (BECs)[1, 2] with laser-like light emission[3] and associated quantum optical phenomena.[4] A wide range of emitters has been investigated with regard to polariton formation and condensation. Organic molecules,[5-7] conjugated polymers[8-10] and fluorescent proteins[11] as well as low dimensional semiconductors such as monolayered transition metal dichalcogenides[12] and single-walled carbon nanotubes[13] have been a special focus over the last decade due to their room-temperature stable excitons and diverse photophysics.

Exciton-polaritons are often created by hybridizing excitons with the fundamental mode of a planar microcavity. Two bright polariton modes, called upper (UP) and lower polariton (LP) are formed as shown schematically in **Figure 1** (center), with the energy gap at the exciton-cavity resonance being the Rabi splitting ($\hbar\Omega$). The energy difference between the exciton and the lowest cavity energy is termed detuning ($\Delta$). In order to achieve polariton condensation the polariton ground state, that is the LP branch zero-momentum $k_\parallel$ state, has to contain a macroscopic population in analogy to BECs.[1] Efficient relaxation of excitations into the LP branch is therefore critical. The relaxation processes can be investigated by injecting polaritons off-resonance *via* exciting the emitter well above the strongly-coupled, lowest excited state. Directly after excitation, internal conversion into the emitter's lowest excited state takes place, populating the so-called exciton reservoir.[3] The reservoir states are associated with high momentum polariton states that can be effectively considered as weakly coupled excitons.[14, 15] Theoretical studies on amorphous organic microcavities, which explicitly model the organic emitter on the molecular level, identify the exciton reservoir as polaritonic dark states.[16, 17] In both scenarios, the reservoir states inherit the character of the underlying molecular excited state and are therefore long lived. Ultrafast spectroscopy studies suggest that the reservoir states still undergo photophysical processes of the weakly coupled emitter.[18, 19] In this picture, UP and LP represent additional decay channels for the excited state of the emitter. Kinetic considerations can be made to determine the fate of the exciton reservoir states.[20] **Figure 1** (center) depicts the three different decay processes of the exciton reservoir that have been proposed as population mechanisms of microcavity exciton-polaritons.[21, 22] Polariton population by scattering of reservoir excitons with acoustic phonons in the case of crystalline solids[15] or molecular translational vibrations in case of amorphous solids[23] is one common mechanism (process i). For molecular emitters, scattering of reservoir excitons with intramolecular vibrations is also possible (vibrationally assisted scattering, VAS) when the Rabi splitting is comparable to the energy of the vibration (process ii).[20, 24] If the scattering processes (i) and (ii) are slow compared to the radiative rate of the reservoir excitons, radiative pumping can take place, i.e., excitons decay directly into the polariton modes (process iii).[25] Understanding the dominant relaxation processes and hence optimizing the employed materials, cavities and experimental conditions is crucial to reach polariton condensation and lasing.

Semiconducting single-walled carbon nanotubes (SWCNTs) have recently emerged as a very interesting material to create not only optically and but also electrically pumped exciton-polaritons in the near infrared (nIR).[13, 26-30] They combine very high ambipolar charge carrier mobilities with large oscillator strength and narrow excitonic absorption and photoluminescence (PL) bands. In addition, cavities with SWCNTs can exhibit Rabi splittings[13] that are comparable to the energy of their longitudinal optical phonons (*e.g.*, the $G^+$ mode).[31] However, up to now no polariton condensation could be demonstrated with SWCNTs and hence understanding their specific polariton population mechanism with respect to their photophysical properties has become crucial.

The photophysics of SWCNT are rather complex (see **Figure 1**, left) and distinct from both inorganic and organic emitters. The geometric and electronic structure of a carbon nanotube can be derived from a rolled-up sheet of graphene and depends directly on the roll-up vector, i.e., the chirality vector (n,m), which determines the diameter and type of nanotubes (metallic or semiconducting). Here we will only consider semiconducting SWCNTs and more specifically (6,5) nanotubes (diameter 0.757 nm), which can be sorted from mixed nanotube raw materials by selective polymer-wrapping in large amounts and with high purity.[32] The SWCNT band structure is that of a one-dimensional semiconductor with van Hove singularities that are nearly symmetrical for holes and electrons. The direct bandgap of nanotubes is inversely proportional to their diameter. The optical transitions of SWCNTs are excitonic with large exciton binding energies (200-400 meV).[33, 34] They are commonly labelled according to the corresponding van Hove singularities with $E_{11}$ and $E_{22}$ and so on. For (6,5) nanotubes the $E_{22}$ transition in thin films is about 2.15 eV (576 nm) and the $E_{11}$ transition is about 1.24 eV (998 nm). Internal relaxation from $E_{22}$ to $E_{11}$ occurs in less than one picosecond[35] and thus emission is only observed from $E_{11}$ in the near-infrared. Furthermore, SWCNTs exhibit a valley structure and spin degeneracy leading to four singlet and twelve triplet excitons. Only one transition is allowed and thus PL from the bright singlet exciton with odd parity and zero center-of-mass momentum, $E_{11}(B)$, is observable.[36] The dark even-parity singlet and all triplet states are energetically below $E_{11}(B)$ while another dark odd-parity singlet but with K-point center-of-mass momentum $E_{11}(K)$ is above.

In addition to a strong and narrow $E_{11}(B)$ emission peak, a series of weak, red-shifted peaks are observed in the emission spectrum of (6,5) SWCNTs, which we will refer to as photoluminescence side bands (PSBs) and are shown in **Figure 1** (left). The $G_1$ transition results from the decay of $E_{11}(B)$ excitons into the ground state under emission of a $G_0$ phonon,[37] whereas the $X_1$ transition originates from momentum-forbidden $E_{11}(K)$ dark excitons, which can only decay radiatively under emission of a $D_0$ phonon.[37, 38] The $Y_1$ transition is believed to be of extrinsic origin and shows tube-to-tube variations.[39] The Ox transition, which we observe for (6,5) SWCNT under ambient conditions, also shows batch-to-batch variations and might be connected to unintentional luminescent oxygen defects.[40]

The photoluminescence quantum yield (PLQY) of (6,5) SWCNTs in dispersions and thin films is relatively low (< 3 %), which is partially attributed to the large number of dark excitons, but more importantly to quenching of the highly mobile excitons at the nanotube ends and non-radiative defects.[41] However, specific defects (variously named luminescent $sp^3$ defects, organic color centers or quantum defects)[42] can trap excitons and serve as radiative recombination sites, thus leading to a substantially increased PLQY (2-8 fold)[43, 44] and also significantly longer fluorescence lifetimes (100 – 500 ps)[45, 46] compared to mobile excitons. The deep optical traps (100-200 meV) lead to a strong red-shift of the emission and facilitate high purity single-photon emission at room temperature.[47, 48] These defects can be created synthetically by arylation,[43, 44] alkylation[49] or covalent oxygen doping.[47] Depending on the binding configuration, two main types of defects have been identified, commonly named $E_{11}^*$ (1.05 eV or 1180 nm) and $E_{11}^{*-}$ (0.95 eV or 1300 nm) for (6,5) nanotubes (see **Figure 1**, right).[50-52] Due to fast diffusive transport of excitons along the nanotubes, only a few luminescent $sp^3$ defects are required to achieve strong emission from them while their contribution to absorption remains negligible.[44] Hence, they might provide a unique way to improve the polariton population by radiative pumping of the LP branch without perturbing the polariton mode structure or creating additional (e.g., middle) polariton modes as usually observed for mixed emitter systems[53] or emitters with pronounced vibronic transitions.[8]

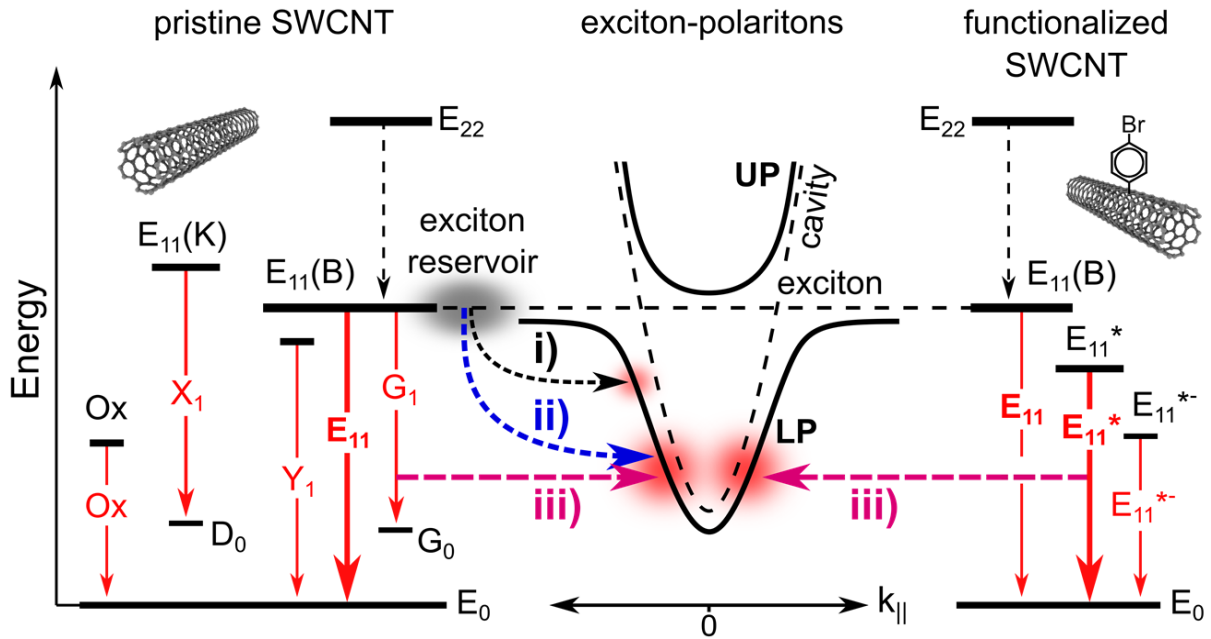

**Figure 1.** Schematic of (6,5) SWCNT energy levels and transitions, and weakly and strongly coupled states in a microcavity. **(Left)** Excitonic states in pristine SWCNTs in the weak coupling regime. The main radiative transition is $E_{11}$ (from the bright $E_{11}(B)$ exciton) followed by weaker, red-shifted transitions ($Y_1$, $X_1$, $G_1$, Ox). $E_{11}(K)$ corresponds to a K-momentum dark exciton. **(Center)** Energy dispersion of SWCNT exciton-polaritons with upper (UP) and lower (LP) polariton mode (solid lines) in relation to the cavity mode and the dispersionless exciton (dashed lines). In case of SWCNTs, the polariton branches might be populated by scattering of reservoir excitons with acoustic phonons (i), with optical phonons (ii) or by radiative pumping (iii). **(Right)** Excitonic states and transitions for a functionalized (6,5) SWCNT. New radiative channels ($E_{11}^{*}$ and $E_{11}^{*-}$) and red-shifted emissions arise from luminescent $sp^3$-defects. Scattering processes from the reservoir excitons of the functionalized SWCNT are omitted for clarity.

Here, we investigate the contributions of different possible population processes to the overall SWCNT exciton-polariton population in optical microcavities with pristine and $sp^3$ defect functionalized (6,5) SWCNTs. By comparing calculated and experimental fluorescence lifetimes of reference films and microcavities with precisely tailored detunings and LP positions, we show that radiative pumping is the predominant polariton population mechanism and that luminescent $sp^3$ defects increase the polariton population up to 6-fold compared to the pristine SWCNTs.

**Results and Discussion**

A typical feature of SWCNT exciton-polaritons in planar microcavities is the noticeable change of emission intensity for different detunings,[13, 26] which should be linked to the polariton relaxation mechanism. Here we show this feature for a metal-clad microcavity with strongly coupled (6,5) SWCNTs embedded in a polymer matrix (**Figure 2**). Thermally evaporated top and bottom gold mirrors with different thicknesses provided broadband reflectivity over the whole range of the InGaAs detector (0.82-1.37 eV) with reasonable quality factors of about 23. The inherent thickness gradient of the spin-coated SWCNT/polymer film from 240 nm to 330 nm enabled the observation and characterization of many different cavity detunings by moving to different positions on the sample. The PL and absorbance spectra of a reference film are given in **Figure S1a-c** (Supporting information). By collecting angle-dependent reflectivity and emission spectra from the cavity in TM polarization *via* Fourier imaging (see Methods section) we could clearly observe the upper and lower polariton modes close to the exciton absorption thus confirming strong coupling (**Figure 2a**). Fitting the polariton modes (TM polarization) to a coupled oscillator model as described in the Methods section gave a Rabi splitting of 100 meV and a detuning of -68 meV.

**Figure 2b** shows the confocally collected PL from the same microcavity as a function of position along the sample and hence film thickness. In this confocal configuration, the polariton emission was integrated over all angles up to 30° (**Supporting Information, Figure S2**). We attribute emission below the $E_{11}$ exciton absorption to the lower polariton branch. The PL from different positions of a (6,5) SWCNT reference film without cavity is provided for comparison (Figure 2b, left). While the emission maximum of the reference remains essentially constant at the transition energy of the exciton ($E_{11}$, 1.227 eV) with intensity fluctuations of about 35 %, the polariton emission exhibits several maxima over the whole detection range with intensity differences of up to 90 %. By scanning along the sample, the film thickness and consequently the detuning of the microcavity is changed. The variation of detuning leads to a red shift of the polariton emission with distinct emission maxima. The occurrence of emission maxima along the LP branch is well-documented for organic exciton-polaritons[5, 54-56,] and indicative of the underlying population mechanism, e.g., vibrationally assisted scattering (VAS). **Figure 2c** reveals that the spectral positions of the observed emission maxima coincide strikingly well with the sideband emission energies of the (6,5) SWCNT PL spectrum, that is, the $Y_1$ (1.205 eV), $X_1$ (1.097 eV), $G_1$ (1.050 eV) and Ox (0.970 eV) sidebands (**Figure 1**). Consequently, these photoluminescence sidebands must play a prominent role in the polariton population of the system. Note that by changing the cavity thickness the transmission at the excitation wavelength also changes, which affects the relative intensities between maxima for different detunings. In this case, it coincides approximately with the emission maximum around $X_1$. We will account for this effect in the population analysis in the last section.

Based on the observed emission pattern in **Figure 2b** we can exclude scattering with acoustic phonons (**Figure 1**, process i) as the underlying population mechanism. Since the scattering rate depends on the phonon density of states (DOS)[23] and the one-dimensional SWCNT acoustic phonons exhibit characteristic van Hove singularities in their DOS,[57,58] a distinct emission pattern should be visible in the polariton PL, which we do not observe (compare **Figure 2b**). Note that even in case of a constant DOS (as sometimes assumed[21]) the vanishing exciton fraction for larger negative detunings (here > 150 meV) renders population by scattering with acoustic phonons rather inefficient. Consequently, we will only consider vibrationally assisted scattering and radiative pumping (processes ii and iii in **Figure 1**) as possible mechanisms for polariton population in this system.

All photoluminescence sidebands could pump the LP radiatively, whereas only the optically active D and G phonons, which are the origins of the $X_1$ and $G_1$ sidebands, may scatter reservoir excitons directly into the LP. Since only the former mechanism is able to account for all observed maxima, we propose that radiative pumping is the dominant population process. The introduction of luminescent *sp³* defects to strongly coupled (6,5) SWCNTs, which should solely pump the polaritons radiatively, is not only a qualitative test and benchmark for this hypothesis (see below) but may also significantly increase the overall polariton population.

Precise control over the detuning of the microcavity is required to explore the impact of detuning and luminescent defects in detail. One of the shortcomings of changing the cavity detuning *via* the film thickness of the emitter layer as shown in **Figure 2** is, that the number of emitters in the cavity also varies and thus the Rabi splitting. We can overcome this issue by creating cavities with uniform dense (6,5) SWCNT films and metal oxide ($AlO_x$) spacer layers with precisely controlled thickness to change the cavity tuning (see Methods section). With this approach we can exclude that the observed increase in emission at more negative detunings arises from an increased number of emitters. Lastly, we ensure that the SWCNT layer is always at the electric field maximum of the cavity and the number of weakly coupled SWCNTs is reduced. Samples with ten different oxide thicknesses were prepared to tune the cavity over the whole SWCNT emission spectrum. To compensate for remaining thickness variations of the SWCNT layers, we employed transfer matrix simulations to predict the LP position for the SWCNT layer thickness of choice, here 80 nm, for each oxide thickness (**Supporting Information, Figure S3**). By locating sample positions with the corresponding LP energy, we were able to control the emitter layer thickness beyond the intrinsic accuracy of the employed spin-coating process for all subsequent experiments.

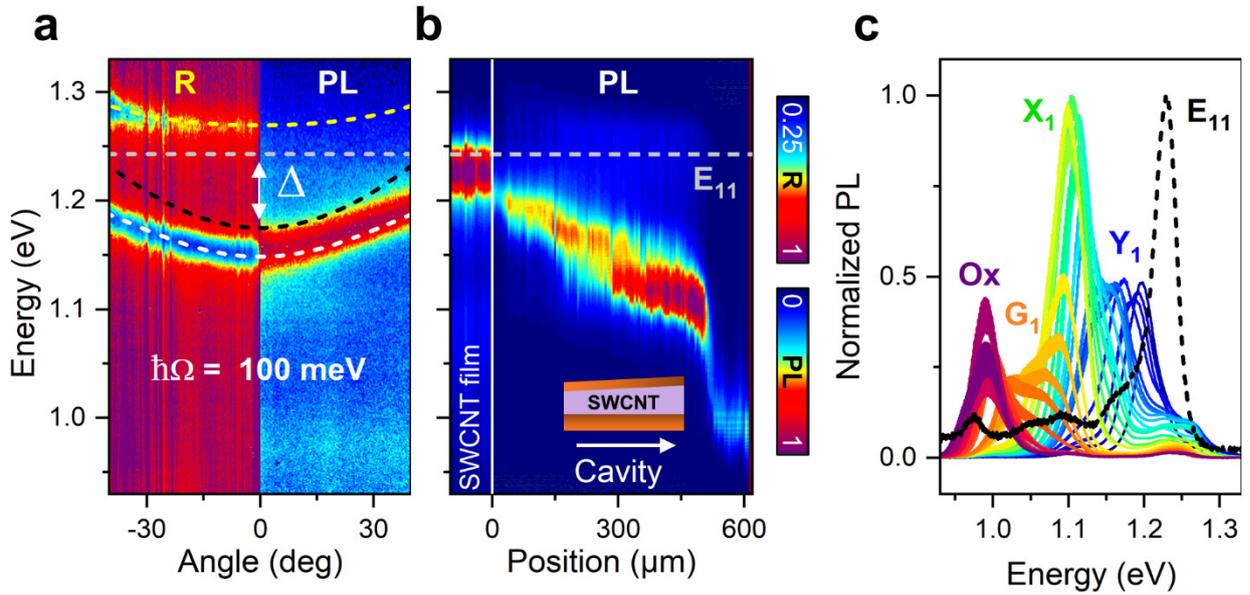

**Figure 2.** PL spectra of a metal-clad microcavity containing a (6,5) SWCNT/polymer film. a) Angle- and spectrally resolved reflectivity (R) and emission (PL) spectra with marked $E_{11}$ energy (gray dashed line), UP (dashed yellow line), cavity (dashed black line) and LP (dashed white line), the detuning is indicated as $\Delta$. b) Angle-integrated PL spectra as function of sample position (*i.e.*, film thickness). The PL of a (6,5) SWCNT film at several positions without a cavity is shown on the left as reference. c) color coded data in (b) for the respective SWCNT sideband transitions ($Y_1$ - blue, $X_1$ - green, $G_1$ – orange and Ox - purple). The PL spectrum of the reference is given as a black dashed line.

To test the radiative pumping hypothesis, we prepared two identical sets of microcavities, one with pristine SWCNTs (**Supporting Information, Figures S4**) and one with 4-bromophenyl-functionalized SWCNTs[44] (**Supporting Information, Figures S5**). The red-shifted emission of the luminescent $sp^3$-defects of the functionalized SWCNTs should exclusively lead to radiative pumping. The functionalization was performed on polymer-sorted (6,5) SWCNTs from the same dispersion batch to exclude processing variations (for a detailed description see Methods section). The degree of functionalization was adjusted to maximize the total SWCNT PLQY (**Supporting Information, Figure S1d**) as it decreases again for very high defect densities.[44] Strong light-matter coupling of both the pristine and functionalized SWCNTs with the various microcavities was characterized by angle-resolved reflectivity and PL spectra. Fits and analysis were based on the coupled oscillator model (see Methods section) and fit results are summarized in **Supporting Information, Figure S6**.

We start by comparing the light-matter coupling of pristine and functionalized SWCNTs in precisely tuned microcavities with oxide spacers. **Figures 3a** and **3c** show the angle-resolved reflectivity and PL spectra of microcavities with each type of SWCNTs tuned to the $E_{11}$ transition. Both samples exhibit splitting into UP and LP modes, which is clear evidence for strong coupling of the $E_{11}$ exciton to the cavity mode. The Rabi splitting is 128 meV for pristine and 106 meV for functionalized SWCNTs. The lower Rabi splitting of the functionalized SWCNTs is the result of a somewhat lower $E_{11}$ absorption (**Supporting Information, Figure S1c**), as the coupling strength scales with the square root of the number of oscillators in the cavity. The cavity emission was studied under non-resonant excitation of the $E_{22}$ transition. We verified that the excitation scheme was suitable for the pristine and functionalized SWCNT filled cavities, respectively, using photoluminescence excitation maps (**Supporting Information, Figure S7**). For both pristine and functionalized SWCNTs, we observe PL only from the LP branch. For $\Delta \approx 0$, the LP emission from the sample with functionalized SWCNTs is about 43 % weaker than that of the pristine SWCNT. We attribute this reduction to the lower $E_{11}$ emission intensity resulting from $E_{11}$ excitons being funnelled to the *sp³* defects.[44] Note that the luminescent *sp³* defects themselves are only weakly coupled as their total number is very small and they do not show measurable absorbance in the SWCNT film around their expected absorption band of 1.086 eV (**Supporting Information, Figure S1c**). Consequently, no splitting at the $E_{11}$* transition energy is observed in reflectivity and thus no additional polariton branches (**Supporting Information, Figure S5**).

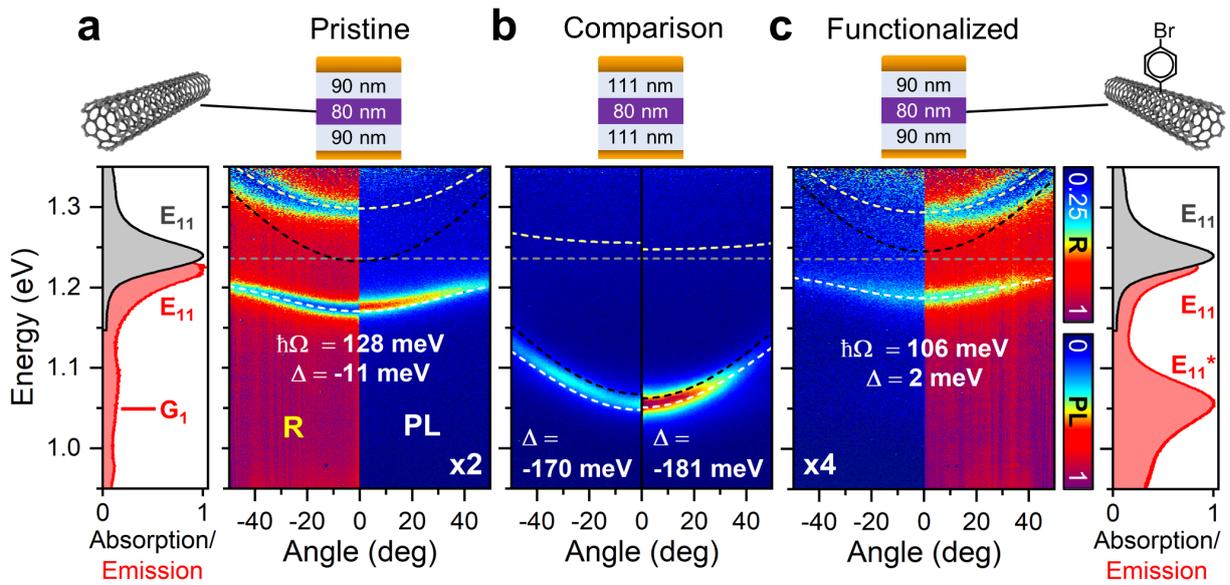

**Figure 3.** Strong coupling with pristine and functionalized SWCNTs. (a) Emission and absorption of a pristine (6,5) SWCNT reference film (left) and angle- and spectrally resolved reflectivity (R) and photoluminescence (PL) of a similar film embedded in a metal-clad microcavity indicating strong coupling. (b) Comparison between angle-resolved PL of two metal-clad cavities with pristine (left) and functionalized SWCNTs (right), tuned to $G_1$ and $E_{11}^*$ transition, respectively. (c) Angle- and spectrally resolved reflectivity and PL of microcavity with functionalized SWCNTs as an active layer and emission and absorption of a functionalized (6,5) SWCNTs reference film (right). The cavity structure is given on top of each data set.

We now turn to a detuning value for which the pristine as well as the functionalized SWCNTs exhibit sideband emission that could radiatively pump the polaritons. **Figure 3b** depicts the angle-resolved PL of microcavities with pristine and functionalized SWCNTs tuned to the $G_1$ and $E_{11}^*$ transition, respectively. The corresponding angle-resolved reflectivity data together with the full coupled oscillator fit results can be found in **Supporting Information, Figures**

**S4-S6**. For Δ ≈ -180 meV the LP emission from microcavity with functionalized SWCNTs is three times stronger than the LP emission from the cavity with pristine SWCNTs. We interpret this enhanced intensity and the spectral position of the LP emission of the cavity with functionalized SWCNTs as indicative of radiative pumping by the $E_{11}^*$ transition, as it is the only mechanism by which this transition can contribute to the polariton population. Note that we assume that all emission from the polariton mode arises from polariton decay. However, the photonic part of the polaritons is an electromagnetic mode and could also act on weakly coupled states by Purcell enhancement. This possibility will be considered and excluded later (see below).

So far, we have obtained qualitative evidence for radiative pumping (**Figure 1**, process iii) of SWCNT exciton-polaritons by introducing luminescent $sp^3$ defects. More direct confirmation of radiative pumping and exclusion of ordinary Purcell enhancement of the PSBs and $sp^3$ defect emission by the polariton mode can be gained by fluorescence decay measurements using time-correlated single-photon counting (TCSPC). Due to the restrictions of the measurement setup, the emission signal was collected from ±20° around $k_\parallel = 0$ (see **Supporting Information, Figure S2**). **Figure 4** (lower panels) depicts the PL decay transients recorded for microcavities with pristine (a) and functionalized SWCNTs (b). The SWCNT layer thickness was kept at 80 nm for all samples and the transients are plotted as a function of the LP energy at $k_\parallel = 0$ for each cavity. The reference spectra of the corresponding (6,5) SWCNT films are shown in the top panels and the contributions of the different PSBs and defect transitions to the spectrum are highlighted as components of a multi-Lorentzian fit.

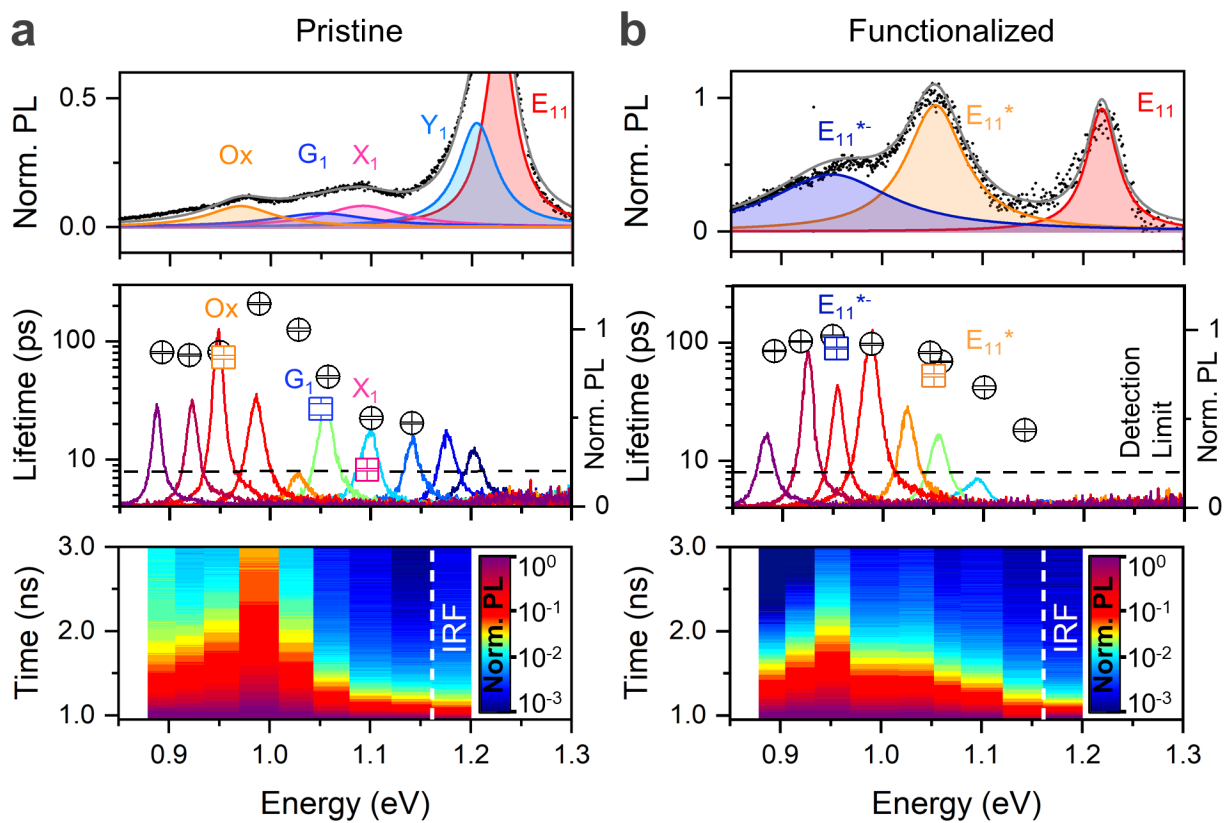

**Figure 4.** a) The top panel shows the multi-Lorentzian fit to the PL of a pristine SWCNT film. The center panel depicts the short lifetime component of the cavity fluorescence decay (black circles) as a function of $k_{\parallel} = 0$ emission energy. The $k_{\parallel} = 0$ emissions of the respective cavities (solid lines) are normalized to the detuning with maximum intensity. The short lifetime components of the SWCNT emission bands without cavity (colored squares) are indicated for comparison. Lower panel: fluorescence decay traces of the cavities and instrument response function (IRF). b) Respective data for functionalized SWCNTs with $E_{11}^*$ and $E_{11}^{*-}$ emission.

All transients, for the microcavities, as well as for the pristine and functionalized reference samples, were well-described by a biexponential decay (for representative histograms and fit results see **Supporting Information, Figures S8** and **S9**). The short lifetime component of individual, pristine SWCNTs has been attributed to the decay of the $E_{11}$ population through radiative and non-radiative channels followed by a slower decay attributed to the redistribution of the exciton population between bright and dark states.[59, 60] Depending on the environment of the nanotubes these lifetimes can be significantly shortened by quenching and even reduced to a monoexponential decay.[60] Indeed, the transients of the $E_{11}$ exciton and $Y_1$ sideband are detection limited and we attribute this fast decay to an increased number of non-radiative decay channels in SWCNT networks compared to individual or freestanding nanotubes. The other PSBs of pristine SWCNTs exhibit values between 8 and 80 ps for the short lifetime component and 100 to 300 ps for the longer lifetime component.

The $E_{11}*$ and $E_{11}*^-$ emission dynamics of functionalized (6,5) SWCNTs with sufficiently low $sp^3$ defect densities, as those employed here, can be considered to be decoupled from the $E_{11}$ exciton dynamics. Here, the short lifetime component is interpreted as the redistribution between trapped bright and dark excitons and the long lifetime component as the subsequent decay through radiative and non-radiative channels.[45, 46] We find 60 to 100 ps for the short and 200 to 300 ps for the long lifetime component of the functionalized (6,5) nanotubes. For both oxide spacer microcavities with pristine and functionalized SWCNTs the fluorescence lifetimes are equal or even slightly longer compared to the corresponding weakly coupled sidebands (**Figures 4a** and **4b**, centre panel and **Supporting Information, Figure S9**). Such similarities between the fluorescence lifetimes of cavities and weakly coupled references were reported previously[14, 61] and interpreted as evidence for radiative pumping by Grant *et al*.[56]

To understand the measured PL decays and lifetimes better, we consider polariton dynamics as well as the Purcell effect. For a kinetic interpretation of the polariton fluorescence decay we make the following assumptions. As can be calculated from the polariton linewidth, the polariton radiative decay in our samples is on the order of a few tens of femtoseconds (**Figure 5**). Hence, we assign the lifetimes observed in the TCSPC experiments as the underlying rate limiting step of the polariton population.[14] Filling the exciton reservoir should occur very rapidly after excitation at 575 nm ($E_{22}$) due to ultrafast conversion from the $E_{22}$ to $E_{11}$ manifold (< 1 ps).[35] Hence, we assume the rate limiting step to be scattering from the exciton reservoir into the polariton states. If the polaritons were radiatively pumped (**Figure 1**, process iii), the observed fluorescence lifetime should be approximately equal to that of the underlying emitter, because a radiative decay of a reservoir exciton must occur prior to polariton population. With this notion we assume that the fraction of weakly coupled radiative decay is not affected significantly by the polaritons or the cavity; an assumption, which we will further discuss in connection with the Purcell effect. If vibrationally assisted scattering (VAS) (**Figure 1**, process ii) occurred in our system, it should lead to a significant reduction of the observed polariton fluorescence lifetime compared to the decay of the weakly coupled reference considering an estimated scattering rate of $(90 - 500 \text{ fs})^{-1}$ for this process (see **Supporting Information** for detailed calculation). In that case, the exciton reservoir would exhibit an additional non-radiative decay channel into the polariton modes and the overall measured fluorescence decay would be shortened substantially.

**Figure 5** shows the experimentally determined short lifetime components of the fluorescence decay of SWCNTs in a microcavity (black diamonds) and the calculated fluorescence lifetimes expected in the VAS limit in the absence of radiative pumping (blue circles). Open symbols represent data for microcavities with pristine SWCNT and closed symbols represent data for microcavities with functionalized SWCNT. For both microcavity and reference we observe a biexponential decay. Within the investigated time frame (10 ns), we can thus exclude a scenario in which the sub-bandgap states transfer population via a non-radiative mechanism with a rate slower than the radiative decay of the weakly coupled SWCNTs. Such an additional decay channel would lead to a noticeably faster decay for the microcavity compared to the reference. We can also exclude a scenario in which the LP decays on the same timescale as the fluorescence from the sub-bandgap states, as this would lead to a triexponential decay as well. Comparing the microcavity lifetimes with the corresponding lifetimes of the pristine and functionalized reference films ($X_1$, $G_1$, Ox, $E_{11}^*$, $E_{11}^{*-}$, colored squares and circles), we find almost identical values, which indicates the absence of VAS and is clear evidence together with the observed biexponential decay for radiative pumping.

Shahnazaryan et al. hypothesized, that the lower polariton may serve as a decay channel for dark excitons in SWCNTs leading to PL quantum yields approaching unity.[62] An activation of dark excitons of the proposed magnitude should drastically shorten the cavity fluorescence lifetime compared to the fluorescence lifetime of the reference, as the dark states could decay via the short-lived LP branch. In steady state cavity PL, the emission (and population) maximum would be observed at the energy of the lowest dark exciton around 1.19 eV. Experimentally, we do not observe either signature, i.e., drastic shortening of the cavity fluorescence lifetime or an emission maximum at the lowest dark exciton energy. Hence, we conclude that activation of dark excitons by polaritons in microcavities based on (6,5) SWCNT networks does not occur or at least not to a detectable degree.

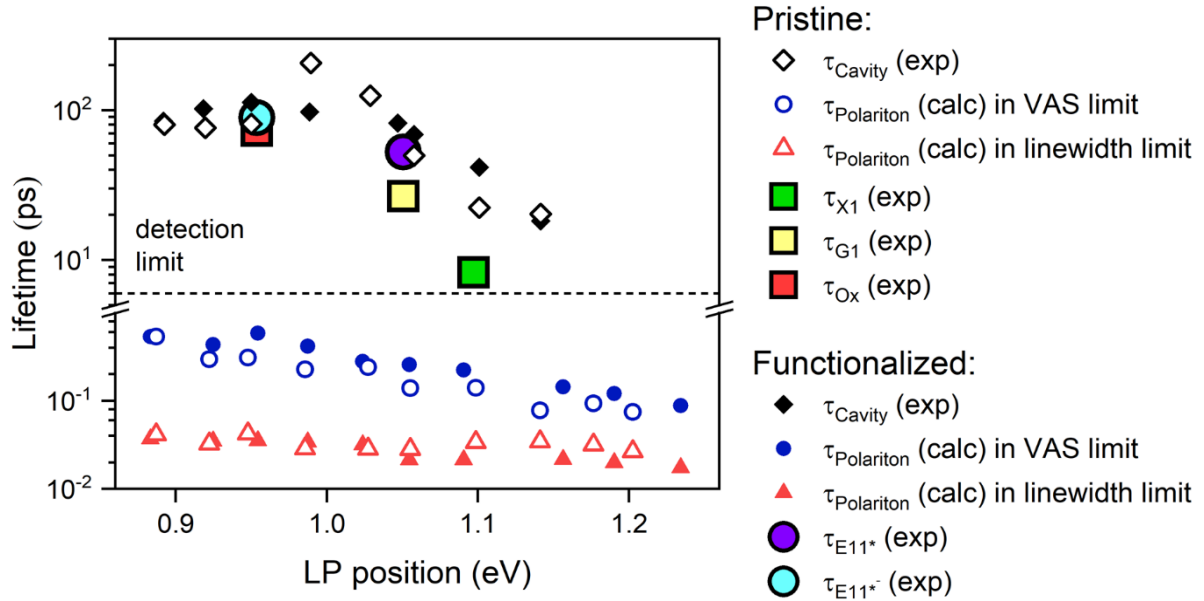

**Figure 5.** Comparison of experimental and calculated fluorescence lifetimes: black diamonds - short lifetime component of the cavity fluorescence decay, blue circles - estimated lifetime for phonon assisted scattering, red triangles - polariton lifetime calculated from the observed LP linewidths. Open symbols correspond to data from pristine SWCNTs and filled symbols to data from functionalized SWCNTs, respectively. Colored large squares (pristine SWCNTs) and circles (functionalized SWCNTs) indicate the short lifetime component of the reference films (no cavity).

So far, we have assumed that the observed emission from the polariton mode only arises from the radiative decay of occupied polariton states. However, the photonic fraction of a polariton is an electromagnetic mode in the classical sense and could enhance a radiative transition, such as a SWCNT PSB, *via* the Purcell effect. In that case, tuning the LP to either the $X_1$ or $E_{11}^*$ transition would result in a 7-fold decrease of the radiative lifetime (see Supporting Information for detailed calculation and **Figures S10** and **S11**). For a moderate non-radiative decay rate, this should lead to a shortening of the fluorescence lifetime, which we do not observe (see **Figure 5**). In fact, we observe slightly longer lifetimes for the microcavities. Note that if the accelerated radiative decay were still small compared to the non-radiative decay, no Purcell enhancement would be observed. This result agrees well with the notion that weakly coupled emission within the cavity is reabsorbed by the polariton states. Simply speaking, if the emission leaked directly out of the cavity, e.g., through the transparent part of the polariton mode, the resulting cavity fluorescence decay would be unavoidably shortened in comparison to the reference (Purcell effect). Based on this argument, we conclude that the observed LP emission solely arises from occupied polariton states.

For an estimate of the relative contributions of radiative pumping by the different PSBs and the $sp^3$ defects to the polariton population, we analyse the polariton population as a function of detuning. The exciton-polariton population $N_P$ is not directly proportional to the PL intensity $I_P$ but to $I_P/\alpha^2$ because the radiative decay increases with the square of the photon fraction $\alpha$ for a given population.[14] **Figure 6** shows the LP population as function of the LP position, calculated from emission averaged over ±1.5° around $k_\parallel = 0$ (see also **Supporting Information, Figure S12**) and corrected for the respective photon fraction determined by the coupled oscillator fit to the corresponding reflectivity data (**Supporting Information, Figures S4** and **S5**). To compare populations at different detunings the data is also corrected for the relative change in excitation efficiency for each cavity structure (**Supporting Information, Figures S13** and **S14**). The uncertainties in **Figure 6** arise from averaging over ±1.5° around $k_\parallel = 0$ and the excitation efficiency correction (**Supporting Information, Figure S14)**. The error calculation is described in the **Supporting Information**. The PL spectra of pristine and functionalized SWCNT films are given as reference.

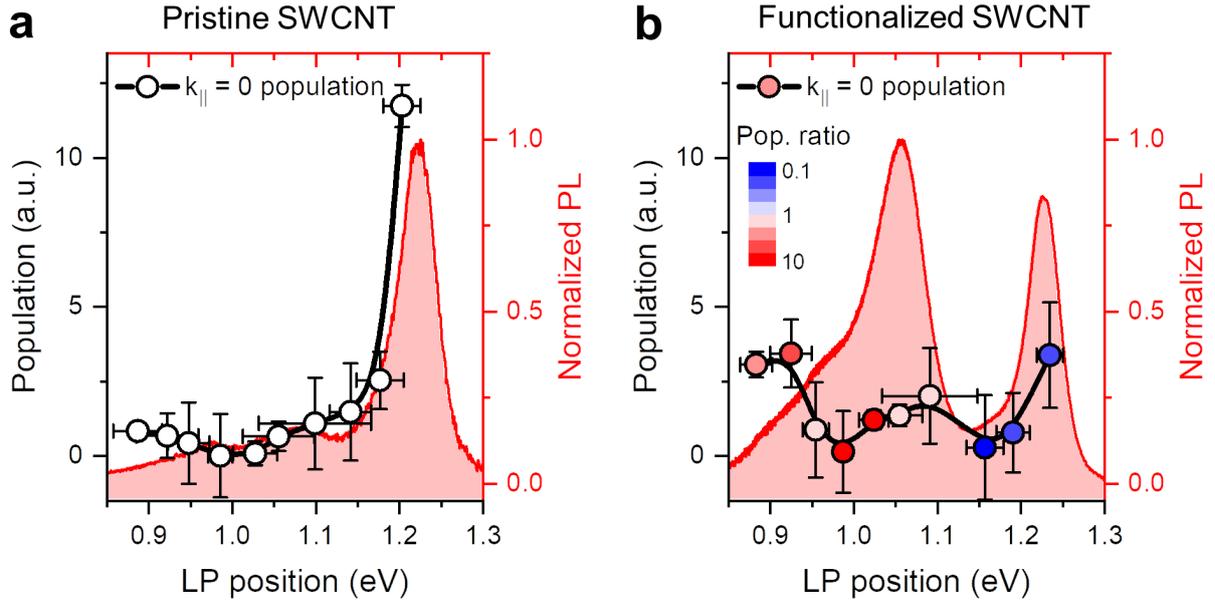

**Figure 6.** Calculated polariton population for SWCNT filled microcavities at $k_\parallel = 0$ as a function of the LP position for (a) pristine SWCNTs and (b) for functionalized SWCNTs. The black solid line is a guide for the eye. For the functionalized SWCNTs, the relative change in population compared to the respective pristine sample is indicated in red for an increase and in blue for a decrease. The PL spectra (red shaded areas) of pristine and functionalized SWCNT films are presented for comparison.

The LP populations in both data sets in **Figure 6** approximately follow the pristine and functionalized SWCNT emission spectra except for a deviation below 1.0 eV. This resemblance strongly suggests that radiative pumping accounts for the majority of the polariton population for the detectable emission angles. The polariton population at $k_\parallel = 0$ of a microcavity with pristine SWCNT is maximized when the LP is tuned to the $E_{11}$ emission (**Figure 6a**). Note that due to the large Rabi splitting of 133 meV this is realized for 85 meV detuning. For the functionalized SWCNTs (**Figure 6b**), the population at 85 meV detuning is reduced to approximately one third of the pristine SWCNTs. This reduction corresponds to the decrease in $E_{11}$ emission for the functionalized reference films compared to pristine SWCNTs and agrees well with population by radiative pumping. Consequently, the polaritons are most likely populated radiatively for small and positive detunings. However, more reliable evidence would require the temporal resolution of the fluorescence decay of the $E_{11}$ transition, which was not possible with the available TCSPC setup.

The LP population of the pristine SWCNTs below 1.0 eV increases slightly and does not follow the emission spectrum. This deviation is even more pronounced for functionalized SWCNTs with luminescent $sp^3$ defects (**Figure 6b**). The underlying further red-shifted defect states ($E_{11}^{*-}$) are believed to have a higher PLQY due to their deeper optical trap depth and increased radiative lifetime,[45] which could be a possible explanation. However, this was not directly corroborated by the fluorescence lifetime data obtained here, and therefore remains elusive.

For the population ratio between functionalized and pristine SWCNTs (**Figure 6b** and **Supporting Information, Figure S15**), we find an increase of the polariton population for all detunings at which the LP overlaps with the $sp^3$ defect emission bands. For LPs overlapping with the $E_{11}$ emission, we find a decrease. This reflects the relative change in the emission spectrum from pristine SWCNT to functionalized SWCNT, as described earlier. For highly emissive polaritons (photon fractions > 98%), we find the highest population, which is 5-fold higher compared to the respective pristine sample (~ 0.92 eV, Fig. 6), while the Rabi splitting is only slightly reduced (~15-25 %). For detunings around 1.0 eV we even find enhancements of about 10-fold (photon fractions > 90%). Additionally, the polariton population depends almost linearly on the $E_{11}^*$ defect emission intensity (tuned by different excitation powers), which further corroborates the notion of radiative pumping (**Supporting Information, Figure S16**).

We conclude that radiative pumping dominates the polariton population of pristine and functionalized SWCNTs. The resulting limitation of the polariton population by the low SWCNT PLQY can be overcome using luminescent $sp^3$ defects. While the total PLQY of functionalized (6,5) nanotubes is at best doubled[44] the polariton population can be increased up

to10-fold. Furthermore, the defect emission can be spectrally tuned by changing the substituents and the binding pattern.[63, 64] Thus, radiative pumping of polaritons by synthetic sidebands constitutes a viable route to decouple the polariton population from the exciton fraction of the polariton state and the phonon DOS of the emitter. In contrast to organic molecules with pronounced vibronic progressions, the oscillator strength of the functionalized nanotubes is not distributed over one or more middle polaritons, owing to the low defect absorption. This distinction makes the observed radiative pumping of SWCNT exciton-polaritons by luminescent *sp³* defects a unique approach to manipulate and increase the polariton population in a resonant cavity. Further angle-resolved fluorescence lifetime measurements on the femtosecond timescale are required to gain more direct insights into the underlying dynamics.

**Conclusion**

With this comprehensive study we have shown that the predominant population mechanism of SWCNT exciton-polaritons is radiative pumping. The polariton fluorescence decay closely resembles the biexponential decay of weakly coupled SWCNT reference samples at various wavelengths, thus excluding vibrationally assisted scattering. The spectral emission shape of both pristine and functionalized SWCNTs can account for the observed polariton population within the investigated range of detunings. The established dominant role of radiative pumping further indicates that polariton population is mainly limited by the low PLQY of SWCNTs. The introduction of luminescent *sp³* defects to the SWCNTs increases the PLQY while only slightly reducing the absorption of the fundamental $E_{11}$ transition and without creating new polariton modes. Hence, microcavities containing functionalized SWCNT exhibit strong coupling with the same polariton branches as those with pristine nanotubes. However, functionalized SWCNTs increase the polariton population at highly emissive detunings (photon fractions > 90%) up to 10-fold. Tuning of the defect emission by changing the substituents and the binding pattern could be further employed to decouple the polariton population from its exciton fraction and tune it to relevant wavelengths. Overall, luminescent $sp^3$-defects constitute a viable and versatile approach toward bright and efficient SWCNT-based polariton devices through radiative pumping.

**Methods**

**Selective Dispersion of (6,5) SWCNTs.** As described previously,[32] (6,5) SWCNTs were selectively extracted from CoMoCAT raw material (Chasm Advanced Materials, SG65i-L58, 0.38 g L$^{-1}$) by shear force mixing (Silverson L2/Air, 10 230 rpm, 72 h) and polymer-wrapping with PFO-BPy (American Dye Source, M$_w$ = 40 kg mol$^{-1}$, 0.5 g L$^{-1}$) in toluene. Aggregates were removed by centrifugation at 60000g (Beckman Coulter Avanti J26XP centrifuge) for 2 × 45 min with intermediate supernatant extraction. The resulting dispersion was split into two parts, one for the pristine SWCNT samples and one for $sp^3$ functionalization.

**SWCNT Functionalization.** Polymer-sorted (6,5) SWCNTs were functionalized with $sp^3$-defects following the previously reported method.[44] Briefly, a toluene solution of 18-crown-6 (99%, Sigma-Aldrich) was added to the as prepared SWCNT dispersion. Subsequently, a solution of 4-bromobenzenediazonium tetrafluoroborate (96 %, Sigma-Aldrich) in acetonitrile was added. The amounts were chosen such that the mixture contained 0.36 mg L$^{-1}$ of (6,5) SWCNT (corresponding to an E$_{11}$ absorbance of 0.2 for 1 cm path length), 7.6 mmol L$^{-1}$ of 18-crown-6 and 0.369 mmol L$^{-1}$ of the diazonium salt in an 80:20 vol-% toluene/acetonitrile solvent mixture. The reaction mixture was left at room temperature in the dark for 16 hours without stirring. Subsequently, the reaction mixture was passed through a PTFE membrane filter (Merck Millipore, JVWP, 0.1 μm pore size) to collect the SWCNTs. The filter cake was washed with acetonitrile and toluene to remove unreacted diazonium salt and excess polymer.

**Film Preparation.** The as-prepared pristine SWCNT dispersion was passed through a PTFE membrane filter (Merck Millipore, JVWP, 0.1 μm pore size) and the filter cake was washed with toluene. Each SWCNT filter cake, pristine and functionalized, contained about 180 μg SWCNT based on absorbance and filtered volume of the respective dispersion. The filter cakes were peeled from the PTFE membrane and transferred into a 5 mL round bottom flask. They were washed three times with toluene at 80 °C for 15 min to ensure complete removal of free wrapping polymer. Subsequently, 0.8 mL of a 2 g L$^{-1}$ PFO-BPy solution in toluene were added and the mixture was sonicated for 1 hour. Afterwards, 0.2 mL toluene were added in 50 μL steps each followed by 15 min of sonication until a homogeneous liquid with a honey-like viscosity was obtained. The final dispersions were spin coated at 2000 rpm on glass substrates (Schott, AF32eco, 300 μm) yielding an average film thickness of about 80 nm with 1.13 wt% SWCNT.

**Microcavity Fabrication.** Prior to deposition, all substrates were cleaned by ultrasonication in acetone and 2-propanol for 10 min, respectively, and UV ozone treatment (Ossila E511, 10 min). A 100 nm thick Au mirror was thermally evaporated onto polished Si substrates with a 2 nm Cr adhesion layer. Subsequently, an AlO$_x$ spacer layer of the respective thickness was deposited by atomic layer deposition (Ultratech, Savannah S100, precursor trimethylaluminum, Strem Chemicals, Inc.) at a temperature of 80 °C. The SWCNT layer was prepared as described above and a second AlO$_x$ spacer was deposited. Thermal evaporation of 40 nm Au as a semi-reflective top mirror completed the cavity.

**Optical Characterization.** All absorption spectra were recorded with a Cary 6000i absorption spectrometer (Varian). For angle-resolved reflectivity measurements, a white light source (Ocean Optics, HL-2000-FHSA) was focused onto the sample by an infinity corrected ×100 nIR objective with 0.85 NA (Olympus, LCPLN100XIR). The resulting spot diameter of ~2 μm defined the investigated area on the sample. For angle-resolved PL measurements, the white light source was replaced with a 640 nm laser diode (Coherent OBIS, 5 mW, continuous wave) and reflected laser light was blocked by a long-pass filter (850 nm cutoff). The reflected/emitted light from the sample was imaged from the back focal plane of the objective onto the entrance slit of a spectrometer (Princeton Instruments IsoPlane SCT 320) using a 4f Fourier imaging system ($f_1$ = 200 mm and $f_2$ = 300 mm). The resulting angle-resolved spectra were recorded by either a 640×512 InGaAs array (Princeton Instruments, NIRvana:640ST) or a 1340×400 Si CCD camera (Princeton Instruments, PIXIS:400) in case of high energy UP modes. A linear polarizer was placed in front of the spectrometer to select between TE and TM polarization.

For PL lifetime measurements, the sample was excited by the spectrally filtered output of a picosecond-pulsed supercontinuum laser source (Fianium WhiteLase SC400) focused by the same objective (Olympus, LCPLN100XIR) and imaged confocally onto an Acton SpectraPro SP2358 spectrograph (grating 150 lines mm$^{-1}$). Scattered laser light was blocked by a dichroic long-pass filter (830 nm cut-off). A liquid nitrogen cooled InGaAs line camera (Princeton Instruments OMA-V) enabled spectral acquisitions to find the desired cavity spectrum. The lifetime measurements were accomplished with a time-correlated single photon counting scheme. The spectrally selected PL emission was focused onto a gated InGaAs/InP avalanche photodiode (Micro Photon Devices) via a ×20 nIR optimized objective (Mitutoyo). Statistics of the arrival times of the photons were acquired with time-correlated single photon counting module (Picoharp 300, Picoquant GmbH). The instrument response function (IRF) was estimated for each sample from the fast, detector-limited PL decay of the (6,5) SWCNTs at the $E_{11}$ transition at 1015 nm.

**Data Analysis and Simulation.** To analyze the experimentally obtained angle-resolved reflectivity and photoluminescence data, the coupled oscillator model was applied.[3] The Hamiltonian in the basis of uncoupled oscillators exciton $|X\rangle = \begin{pmatrix}1\\0\end{pmatrix}$ and photon $|C\rangle = \begin{pmatrix}0\\1\end{pmatrix}$ is

$$\mathcal{H} = \begin{pmatrix} E_X - i\hbar\Gamma_X & V_A \\ V_A & E_C - i\hbar\Gamma_C \end{pmatrix} \quad (1)$$

, where $E_x$ is the energy of the $E_{11}$ excitonic transition and $\hbar\Gamma_X$ is the half width at half maximum (HWHM) for the homogeneously broadened line. The energy dispersion of the cavity is given by

$$E_C(\theta) = E_0 \left(1 - \left(sin(\theta)/n_{eff}\right)^2\right)^{-1/2} \quad (2)$$

for a cavity tuned to $E_0(\theta) = E_X + \Delta$, with $\Delta$ being the cavity detuning and $\hbar\Gamma_C$ the HWHM of the cavity mode. The coupling potential $V_A$ of the two oscillators is related to the Rabi splitting at $E_C = E_X$ by $V_A = \sqrt{\hbar\Omega^2 + (i\hbar\Gamma_C - i\hbar\Gamma_X)^2}$.

With the eigenvalues of the Hamiltonian

$$E_{UP/LP} = \frac{1}{2}(E_X - i\hbar\Gamma_X + E_C - i\hbar\Gamma_C) \pm \frac{1}{2}\sqrt{V_A^2 + 0.25(E_X - i\hbar\Gamma_X - E_C + i\hbar\Gamma_C)^2} \quad (3)$$

in the basis of the coupled oscillators $|UP\rangle$ and $|LP\rangle$ the experimental UP and LP dispersion can be fitted. As initial value for the effective refractive index $n_{eff}$ the geometric mean of host polymer and spacer refractive indices at 998 nm was used (2.36 for TE and 2.28 for TM). The photonic (excitonic) fractions $\alpha$ ($\beta$) of the new eigenstates (Hopfield coefficients) were calculated by their projection onto the uncoupled cavity and exciton modes, that is, $\alpha_{UP} = |\langle UP|C\rangle|^2$ and $\beta_{UP} = |\langle UP|X\rangle|^2$ for UP and likewise for LP. Transfer-matrix simulations were performed as described previously.[13] For the given SWCNT concentration of 1.13 wt%, the sample reflectivity was simulated for different thicknesses (see **Figure S3, Supporting Information**). With this approach, the desired SWCNT layer thickness on a given sample was found experimentally by matching the lower polariton energetic position to that of the simulation.

ASSOCIATED CONTENT

SWCNT film and dispersion characterization, analysis of angle-integrated polariton emission spectra, thickness determination by transfer matrix simulation, coupled oscillator fit results and angle-dependent spectra, cavity PLE spectra, fluorescence lifetime measurements, estimate of mean exciton distance, polariton fluorescence decay lifetime in the limit of VAS, absence of Purcell effect and implications for radiative pumping, polariton emission at $k_\parallel = 0$ as a function of detuning, excitation efficiency correction for population analysis, LP population as a function of defect emission (PDF).

AUTHOR INFORMATION

**Corresponding Author**

*E-mail: zaumseil@uni-heidelberg.de

**ORCID**

Jana Zaumseil: 0000-0002-2048-217X

ACKNOWLEDGMENT

The authors thank A. Mischok for valuable discussions. This research was funded by the Volkswagenstiftung (Grant No. 93404). F.J.B. acknowledges funding by the European Research Council (ERC) under the European Union's Horizon 2020 research and innovation programme (Grant agreement No. 817494 "TRIFECTs").

# Supporting Information

# Population of Exciton-Polaritons *via* Luminescent *sp³* Defects in Single-Walled Carbon Nanotubes

*Jan M. Lüttgens, Felix J. Berger, Jana Zaumseil\**

Institute for Physical Chemistry and Centre for Advanced Materials, Universität Heidelberg, D-69120 Heidelberg, Germany

\*zaumseil@uni-heidelberg.de

# SWCNT film and dispersion characterization

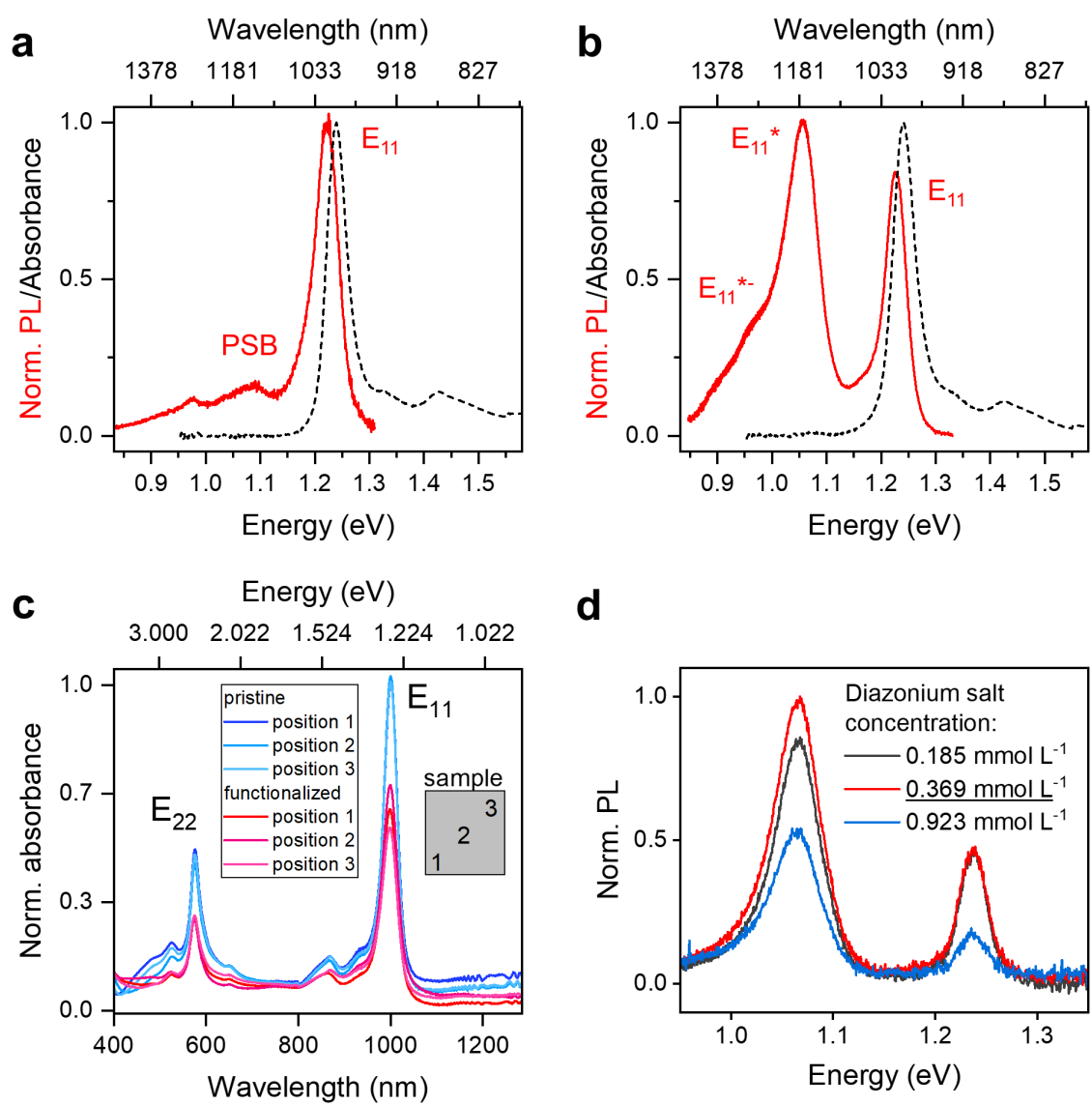

**Figure S1.** PL spectra (red solid line) of (a) pristine (normalized to $E_{11}$) and (b) functionalized (normalized to $E_{11}$*) (6,5) SWCNT films with PFO-BPy as the matrix polymer on glass. Absorbance spectra of both films are shown for reference (black dashed line). c) Absorbance spectra for the same films at three different sample positions corrected for film thickness and normalized to the $E_{11}$ absorbance of the pristine SWCNTs. d) PL spectra of functionalized (6,5) SWCNT dispersions in toluene for three different diazonium salt concentrations. The spectra are normalized to the $E_{11}$* emission of the sample with 0.369 mmol L$^{-1}$ (red) of diazonium salt in the reaction mixture, which was used for cavity fabrication in this study. Note that higher diazonium salt concentrations lead to a decrease in overall photoluminescence quantum yield (blue curve) as described in detail by Berger *et al.*[1]

## Analysis of angle-integrated polariton emission spectra

Due to the geometric restrictions of the fluorescence mapping and TCSPC setup, the polariton emission cannot be angularly resolved and was collected confocally. The angle-integrated (confocal) PL measured with this setup (**Figure S2a**) agrees well with the sum over the angle-resolved PL from -30° to 30° (**Figure S2b**). Accordingly, the data in Figure 2 corresponds to an integration over these angles. For the fluorescence lifetime measurements, only emission at the wavelengths corresponding to $k_\parallel = 0$ was collected. The maximum range of angles that contribute to this emission is ±20° for cavity detunings up to -33 meV based on the FWHM (full width at half maximum) (**Figures S2b and S2c**). Note that for detunings closer to resonance, the curvature around $k_\parallel = 0$ decreases, leading to an increased contribution from larger angles.

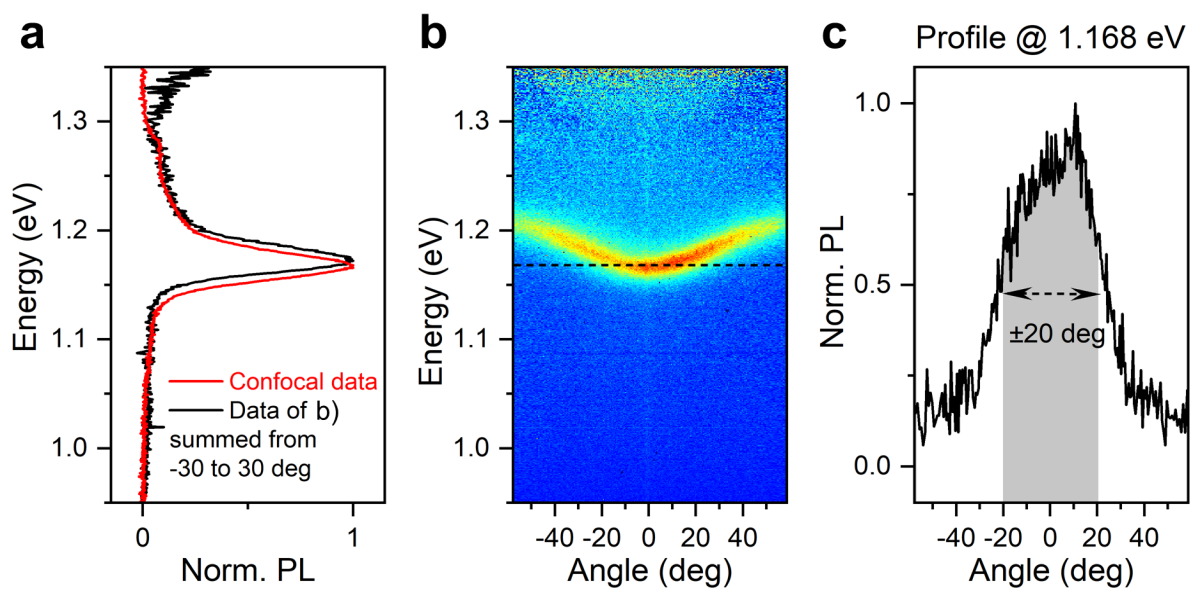

**Figure S2.** a) Normalized PL of the same sample measured confocally at a position with similar detuning (red line) and the sum of the data in (b) from −30° to 30° for comparison (black line). b) Angle-resolved PL of a SWCNT microcavity in the strong coupling regime c) Intensity profile of data in (b) at 1.168 eV.

## Thickness determination by transfer matrix simulation

Spin-coated films exhibit inherent thickness gradients, which result in height differences of several tens of nanometers over a large sample area. The films prepared in this work varied by about 40 nm in thickness from the sample center to the edge. To account for this variation, we simulated the cavity reflectivity using transfer matrix simulations as described previously.[2]

The oxide spacer and gold mirror thicknesses can be assumed to be constant for one sample. All observed changes in the polariton modes are produced by the variation of the SWCNT layer thickness (assuming the absence of large defects or aggregates). Accordingly, the polariton mode position is indicative of the layer thickness. By simulating the reflectivity spectra for the employed SWCNT concentration for a given thickness, an area with the desired layer thickness can be identified on the sample by matching the lower polariton energetic position with that of the simulation.

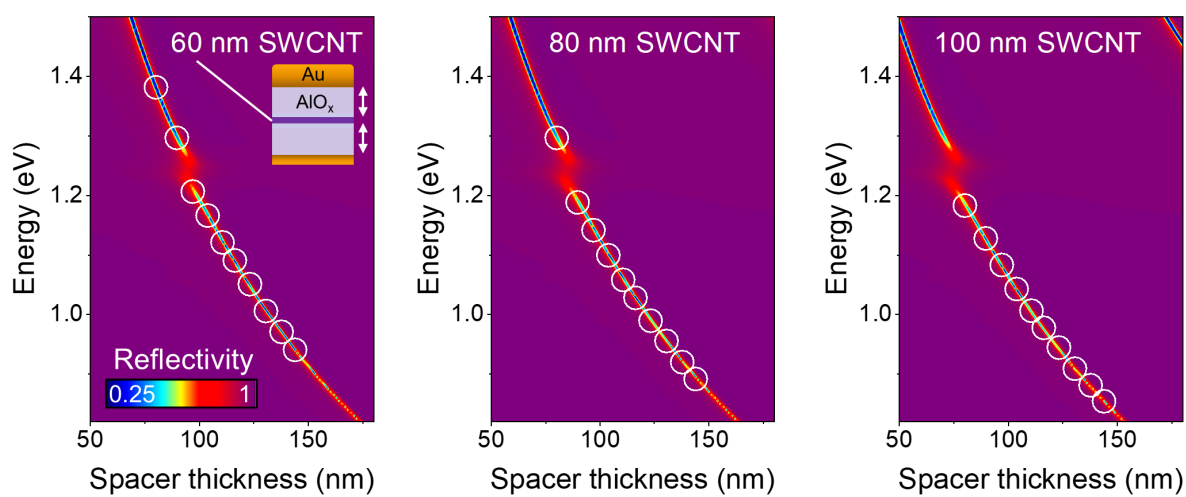

**Figure S3.** Transfer matrix simulations of the reflectivity of the cavity stack as a function of oxide spacer thickness for SWCNT films with 60, 80 and 100 nm thickness. The thickness of one oxide spacer is given on the x-axis. The cavity stack is shown as an inset. White circles indicate the polariton mode energy (UP for positive, LP for negative detunings) for 10 of the produced oxide spacer thicknesses.

## Coupled oscillator fit results and angle-dependent spectra

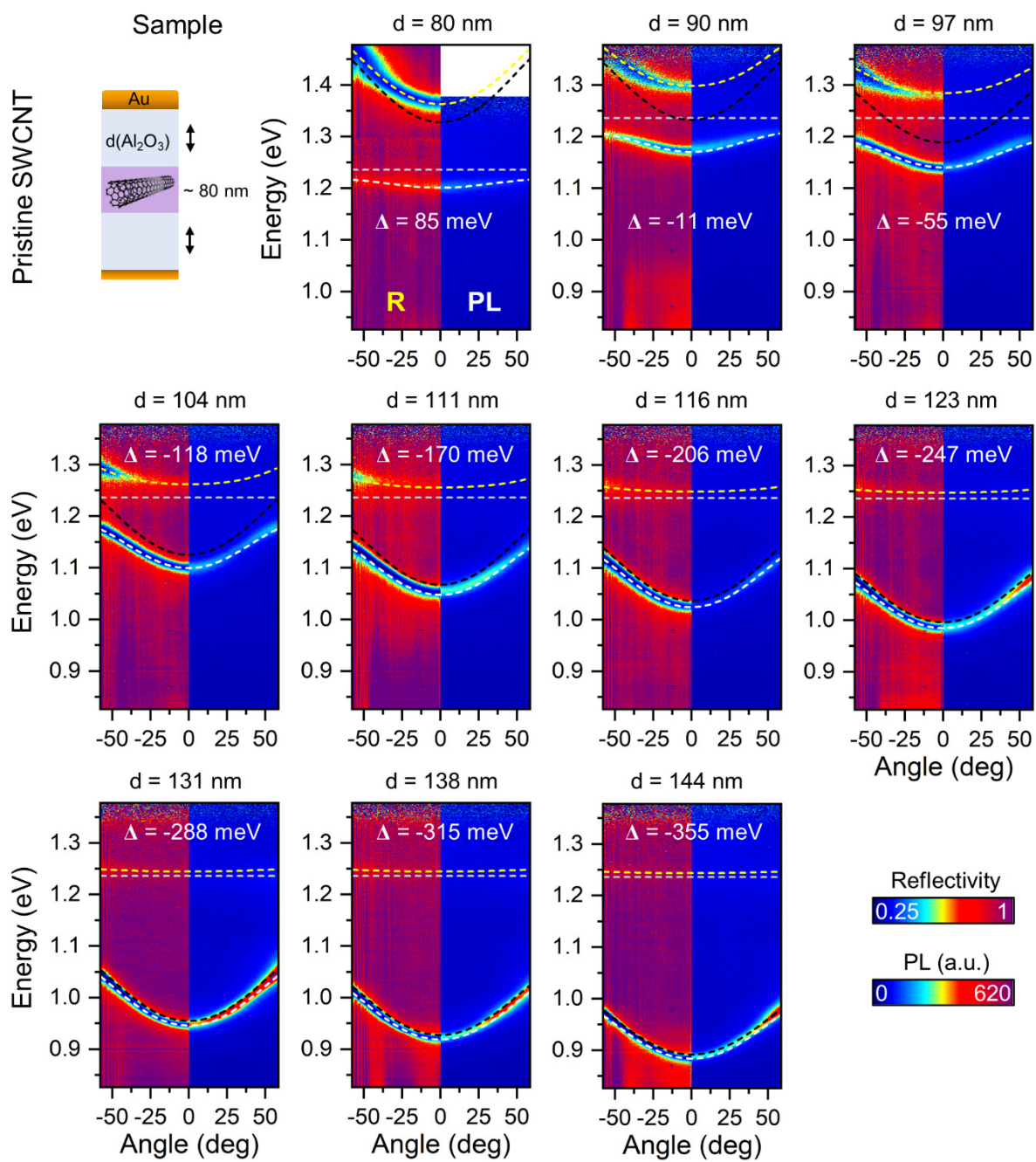

**Figure S4.** Angle-resolved reflectivity (R) and photoluminescence (PL) spectra for a cavity with an 80 nm thick film of pristine SWCNTs for different oxide spacer thicknesses. The fitted modes are given as dashed lines: UP (yellow), cavity (black) and LP (white). The energy of the $E_{11}$ exciton is indicated by a dashed grey line. The oxide spacer thickness and the detuning obtained from the fits are presented for each data set. For the sample with 85 meV detuning, the upper polariton was recorded with a Si CCD camera (Princeton Instruments, PIXIS:400).

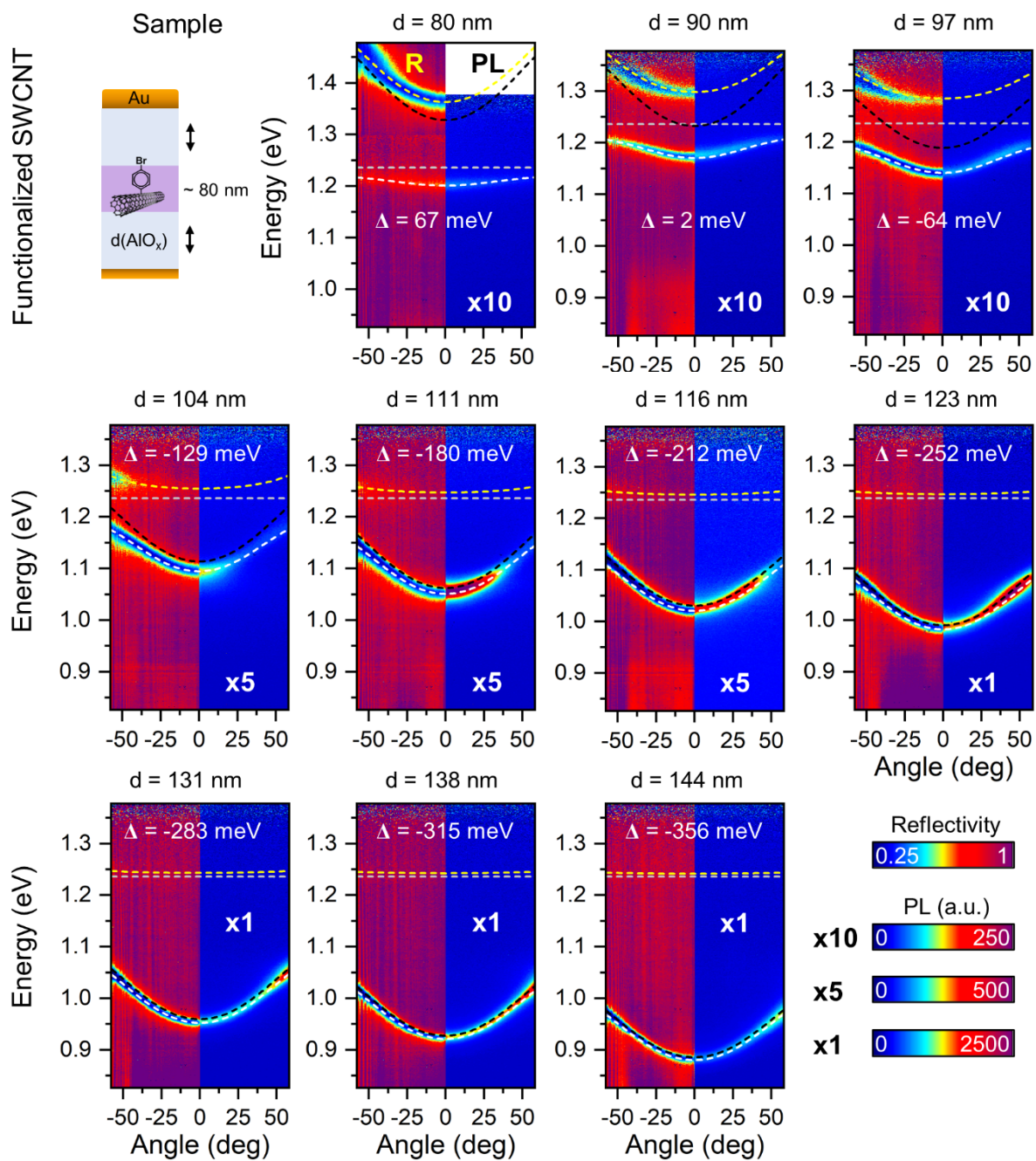

**Figure S5.** Angle-resolved reflectivity (R) and photoluminescence (PL) spectra for a cavity comprising an 80 nm thick film of functionalized SWCNTs for different oxide spacer thicknesses. The fitted modes are given as dashed lines: UP (yellow), cavity (black) and LP (white). The energy of the $E_{11}$ exciton is indicated by a dashed grey line. The oxide spacer thickness and the detuning resulting from the fit are presented for each data set. For detunings of 67 meV to -212 meV the color scale was adapted to make the LP emission visible. For the sample with 67 meV detuning, the upper polariton was recorded with a Si CCD camera (Princeton Instruments, PIXIS:400).

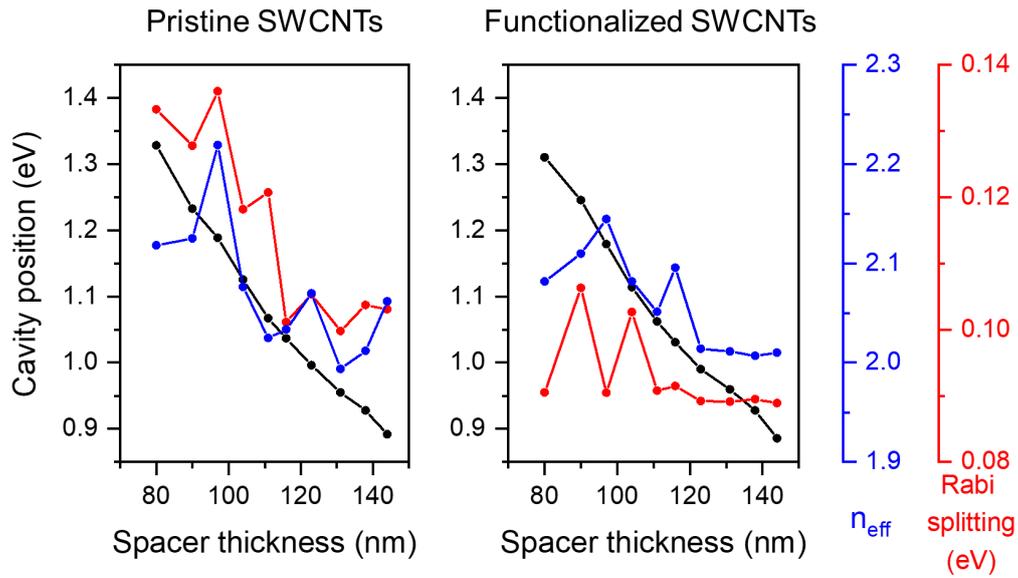

**Figure S6.** Results of coupled oscillator fits to the polariton modes recorded in reflectivity (TM polarization) for pristine and functionalized (6,5) SWCNTs. The results are sorted by spacer thickness of the cavity. The SWCNT layer thickness is always 80 nm. The fitted modes and the corresponding reflectivity data are shown in **Figure S4** and **S5**.

# Cavity PLE spectra

Cavity photoluminescence-excitation (PLE) maps were recorded by focusing the spectrally filtered output of a picosecond-pulsed supercontinuum laser source (Fianium WhiteLase SC400, ~6 ps pulse width, 20 MHz repetition rate, ~20 μJ cm$^{-2}$ pulse energy) through a ×100 nIR-optimized objective (N.A. 0.85, Olympus) onto the sample. The angle-integrated emission spectra were recorded for excitation wavelengths from 520 to 880 nm using a spectrograph (Acton SpectraPro SP2358) and an InGaAs line camera (Princeton Instruments OMA V). The recorded data was corrected for the wavelength dependence of the laser output power and the detection sensitivity.

**Figure S7** shows the PLE maps for a microcavity with a pristine (a) and a functionalized (b) SWCNT layer. Both cavities are tuned to a PL transition around ~1130 nm, corresponding to $X_1$ in the case of the pristine SWCNT and to the blue flank of the $E_{11}^*$ in case of the functionalized SWCNT. For both samples the PLE emission maximum is observed at this wavelength. The emission maxima correlate with 575 nm and 860 nm in excitation, thus corresponding to the SWCNT $E_{22}$ absorption peak and the phonon sideband of the $E_{11}$ absorption, respectively. This correlation confirms that the origin of the cavity emission is the excitation of the (6,5) SWCNT at higher excited states for both the pristine and the functionalized SWCNT, respectively. The $E_{11}$ emission in both cases is only faintly visible, resulting from a weakly coupled fraction of $E_{11}$ excitons. This observation agrees well with a strongly coupled $E_{11}$ transition. The lower intensity of the 860 nm excitation peak in comparison to the references (**Figures S7c** and **S7d**) is attributed to the onset of the gold mirror reflectivity. The lower intensity of the $E_{11}^*$ compared to the $E_{11}$ transition in **Figure S7d** is due to the high peak power of the picosecond-pulsed laser, which already starts to saturate the $E_{11}^*$ transition.[1]

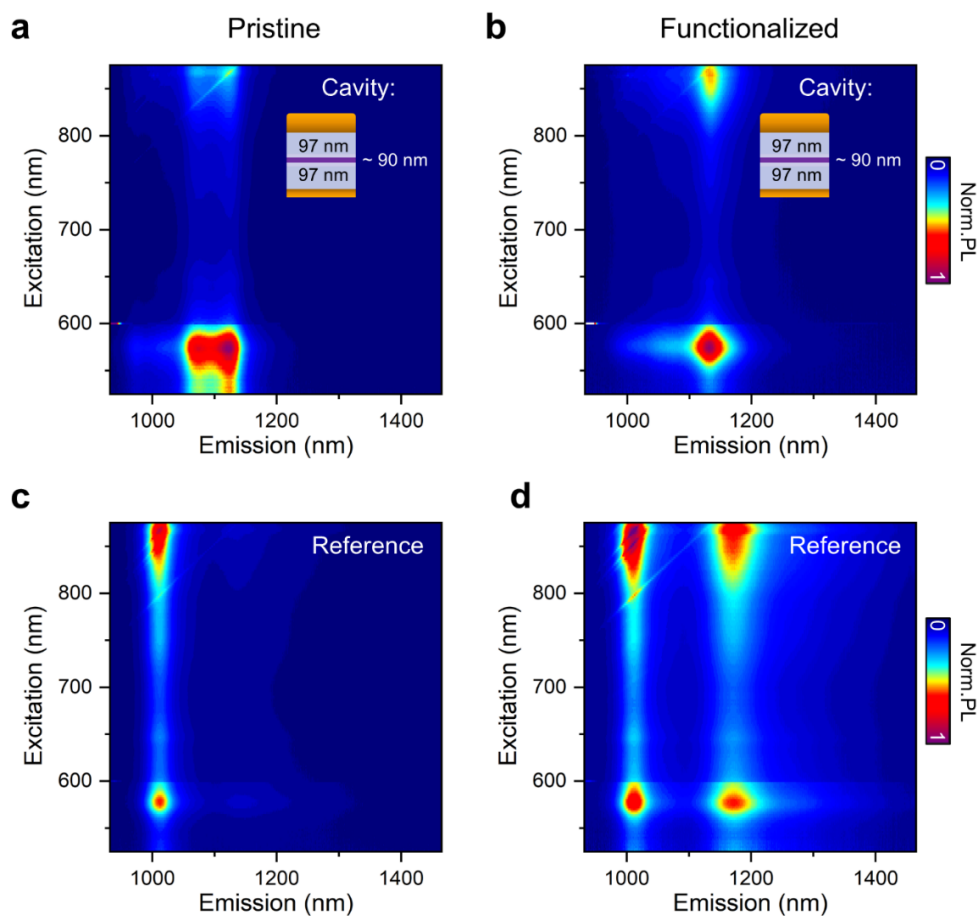

**Figure S7.** PLE maps of microcavities filled with pristine (a) and functionalized (b) SWCNT tuned to the $X_1$ and the $E_{11}^*$ transition, respectively, and the corresponding pristine (c) and functionalized (d) SWCNT reference films on glass. The cavity structure is shown in the inset. Note that the lower intensity of the $E_{11}^*$ compared to the $E_{11}$ transition in (d) is due to the high peak power of the picosecond-pulsed laser, which already starts to saturate the $E_{11}^*$ defects.

**Fluorescence lifetime measurements**

The cavity fluorescence decay dynamics were investigated using time-correlated single photon counting (TCSPC). The sample was mounted onto a microscope stage and the excitation laser (575 nm, ~6 ps pulse width, 20 MHz repetition rate, ~20 µJ cm$^{-2}$ pulse energy) was focused through a ×100 nIR-optimized objective (N.A. 0.85, Olympus). As described in the previous section, for each lifetime measurement an angle-integrated emission spectrum was recorded using a spectrograph (Acton SpectraPro SP2358) and an InGaAs line camera (Princeton Instruments OMA-V) to select the sample position of interest. The cavity emission wavelength at $k_{\parallel} = 0$ was selected and focused onto a gated InGaAs/InP avalanche photodiode (Micro Photon Devices) through a ×20 nIR-optimized objective (Mitutoyo). Statistics of the arrival times of the photons were acquired with a TCSPC module (Picoharp 300, Picoquant GmbH). The same procedure was used to measure the fluorescence decay of the pristine and functionalized SWCNT reference films of the respective PL bands. The instrument response function (IRF) was estimated for each sample from the fast, detector-limited PL decay of the (6,5) SWCNTs at the $E_{11}$ transition at 1015 nm. All decay curves were fitted to a biexponential model in a reconvolution procedure correcting for the IRF. The detection limit based on the IRF FWHM of ~80 ps using this procedure was 8 ps.

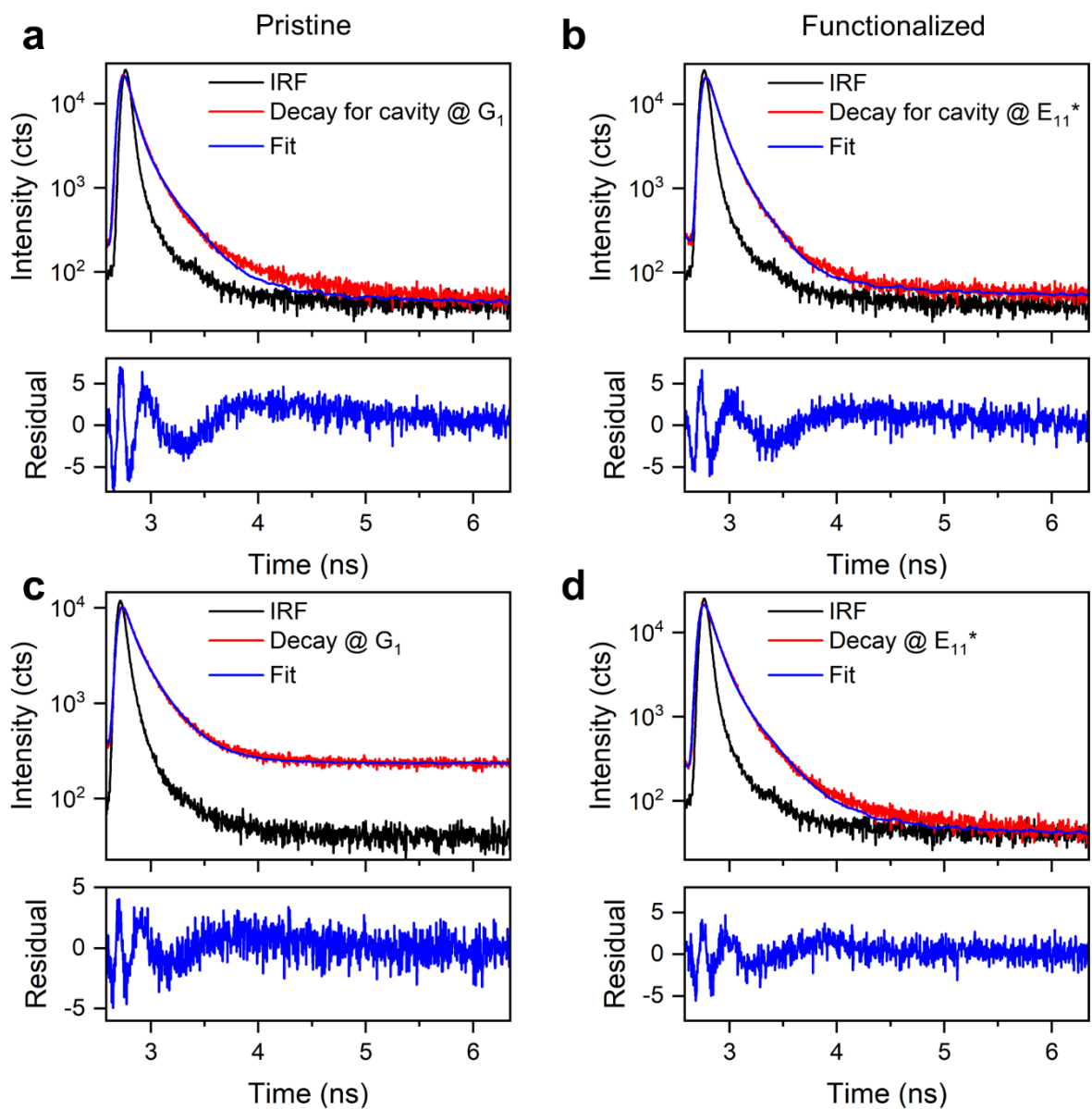

**Figure S8.** TCSPC histograms showing the fluorescence decay (red curve, upper plots) for a cavity with pristine SWCNTs tuned to the $G_1$ band (a), a cavity with functionalized SWCNTs tuned to the $E_{11}^*$ band (b). The respective decays of reference films on glass are depicted in (c) and (d). The biexponential fit (blue curve) was determined with a reconvolution procedure considering the IRF (black curve). Residuals of the respective fits to the decays are given in the lower plots. The spiking of the residual at early times results from the reconvolution.

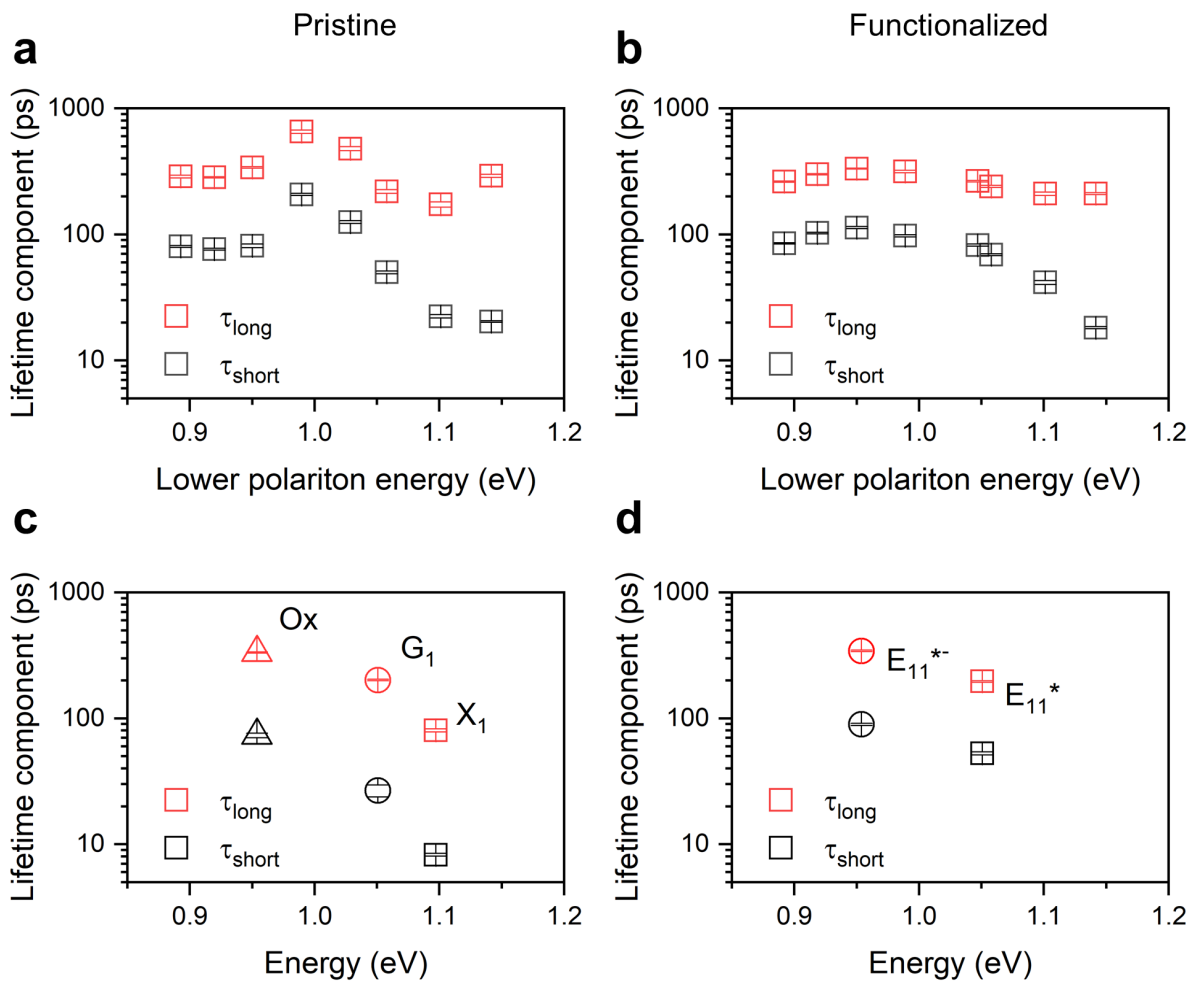

**Figure S9.** Long and short lifetime components of the cavity fluorescence decay as a function of lower polariton energy for pristine (a) and functionalized (b) SWCNTs. (c, d) Short and long lifetimes for the respective reference films as a function of emission energy. The error bars represent the error of the biexponential fit and reconvolution.

## Estimate of mean exciton distance

For the lifetime measurements, samples were excited at 575 nm ($E_{22}$ transition) with pulsed laser light (20 MHz, ~6 ps pulse width). The average laser power was ~100 µW corresponding to $1.4 \times 10^7$ photons per pulse. The film absorbance at the $E_{22}$ transition was determined to be ~0.15 for the pristine and ~0.1 for the functionalized samples, corresponding to an absorption of ~29 % and ~21 % of incident photons, respectively. For simplicity, we will use the average value in the following. The $E_{22}$ to $E_{11}$ conversion efficiency was estimated to be around 91 %.[3] However, complete conversion takes about 5 ps such that the generated $E_{11}$ excitons can already diffuse to quenching sites or *sp³* defects.[4] We can estimate the fraction of freely diffusing excitons, that are not quenched or trapped, *via* the absolute PL quantum yield of the pristine film, which is around 0.2 %.[1] With this value we calculate $6.4 \cdot 10^3$ excitons per pulse. Considering that the beam was focused on a ~2 µm diameter spot and the presented lifetime data was recorded for an 80 nm thick SWCNT-polymer layer, the excited volume was about $10^{-12}$ cm³. This yields an exciton density of $6.4 \cdot 10^{15}$ cm⁻³ per pulse. As the SWCNT are dispersed homogeneously in the polymer matrix, we can approximate the mean exciton distance with Wigner-Seitz radius yielding 155 nm. The error of the mean exciton distance mainly arises from the uncertainty of the laser pulse power and excitation spot diameter. It enters through the estimation of the exciton density. We estimate that the uncertainty in these quantities might lead to a variation of the rate by a factor of three at maximum.

## Polariton fluorescence decay lifetime in the limit of VAS

We estimate the scattering rate $W_{\text{VAS}}$ for vibrationally assisted scattering (VAS) in the adiabatic approximation by:[5]

$$W_{\text{VAS}} = \left(\frac{a}{L_c}\right)^3 \frac{\pi^2 g^2 (\Omega/2)^2 E_c(k_0^2)}{\hbar E_c^2(0)} = 1/\tau_{\text{VAS}}$$

We replace the mean distance between molecules ($a$) by the mean distance of excitons in the SWCNT, which we estimated to be 155 nm in the TCSPC experiments (see previous section). For metal cavity mirrors we can approximate the cavity length as $L_c = \frac{\lambda_{\text{res}}}{2n_{\text{eff}}}$, where $\lambda_{\text{res}}$ is the resonant wavelength and $n_{\text{eff}}$ is the cavity effective refractive index. The dimensionless exciton-phonon coupling constant $g$ was experimentally determined for the G phonon as 0.9.[6] Note that we neglect the radial breathing mode (RBM) here, as it is not able to transfer relative momentum.[7] The coupling constants for the remaining SWCNT optical phonons were determined theoretically and have comparable magnitudes to $g_G$.[7] For simplicity, we will assume the existence of an optical phonon for each investigated cavity detuning using $g_G$ as the coupling constant. $E_c(0)$ is the cavity resonance at $k_{||} = 0$ and $E_c(k_0^2)$ is the cavity energy for the position $k_{||} = k_0$, where the energy of the optically active phonon $E_{\text{vib}}$ matches the energy difference $\Delta E = E_X - E_{\text{LP}}(k_0^2)$ between exciton reservoir and LP. All cavity parameters as well as the Rabi splitting $\hbar\Omega$, are taken from the coupled oscillator model fitted to the experimental angle-dependent reflectivity data of the respective sample. With these values we calculate $\tau_{\text{VAS}} \approx 90 - 500$ fs depending on the cavity detuning. The results are shown in **Figure 5**. Given such a high scattering rate, we would expect the fluorescence decay to be detection limited.

## Absence of Purcell effect and implications for radiative pumping

To understand the nature of the observed emission enhancement by radiative pumping, it is helpful to conduct the following thought experiment: As reported previously,[8-11] the exciton reservoir still undergoes decay processes as in the weak coupling regime because only a small fraction of excitons (1-30 %)[12] is coherently coupled to the light field. The polariton modes rather represent additional states, into which the reservoir population can decay. Starting from this, we consider a radiative transition of the exciton reservoir to the ground state and the resulting photon as weakly coupled. Secondly, we consider the photonic part of the polariton modes as electromagnetic modes in the classical sense. If the polariton is tuned to the energy of the photon, this should lead to Purcell enhancement of the underlying transition. For the condition that the SWCNT layer lies at the anti-node of the electric field created by the polariton (**Figure S10**), we can estimate the Purcell factor $F_P^{2D}$ by:[13, 14]

$$F_P^{2D} = \frac{1}{4\pi} \frac{\lambda_{\text{res}}}{n_{\text{eff}} L_{\text{c}}} Q$$

where the resonant wavelength is $\lambda_{\text{res}}$ (1.1 meV), here tuned to $X_1$ (1.092 meV). The cavity effective refractive index $n_{\text{eff}}$ (2.08) and the quality factor of $Q_{\text{exp}}$ (44.9) were taken from the experimental reflectivity data of a strongly-coupled microcavity with pristine SWCNT tuned to $X_1$ (**Figure S4**, $\Delta = -118$ meV). The cavity length $L_{\text{c}}$ is $\frac{\lambda_{\text{res}}}{2 n_{\text{eff}}} = 271$ nm (see **Figure S10**). With these values we calculate $F_P^{2D} = 7.2$. This Purcell factor is equivalent to a sevenfold decrease of the radiative lifetime in the cavity compared to free space. Such a considerable increase of the radiative decay should be observable in the fluorescence decay. Note that the non-radiative decay is not altered by the cavity, accordingly any enhancement due to the Purcell effect must also affect the fluorescence lifetime. If the enhancement of the radiative decay is small compared to the intrinsic non-radiative decay, also no significant increase of the LP emission should be observable. **Figure S11** shows TCSPC time traces of a cavity with pristine and functionalized SWCNTs and their respective reference films measured at the same energy. Note that no change of fluorescence lifetime is found. So far, we could not measure Purcell enhancement for any of our strongly coupled cavities. Based on this observation, we consider the weakly coupled photons as being absorbed by the polaritons and the resulting emission as polaritonic. In a potential fluorescence up-conversion experiment (currently not available) the fluorescence decay of the reference should be faster than that of the polaritons, as the photons will remain within the sample for the duration of the lifetime of the polaritons.

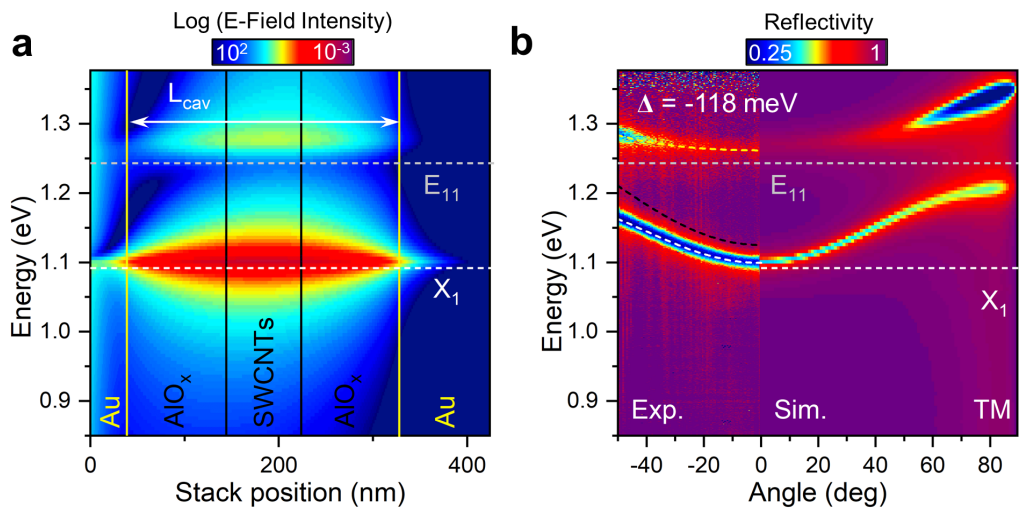

**Figure S10.** (a) Electric field intensity of the photonic part of the polariton at $k_\parallel = 0$ in the microcavity of **Figure S4** ($\Delta = -118$ meV) from transfer matrix calculations. The different materials of the device stack and the relevant transitions of pristine SWCNTs are indicated. (b) Experimental and simulated angle-dependent reflectivity of the same sample. The photon fraction of the LP at $k_\parallel = 0$ is 85 % (based on coupled oscillator fit to experimental reflectivity).

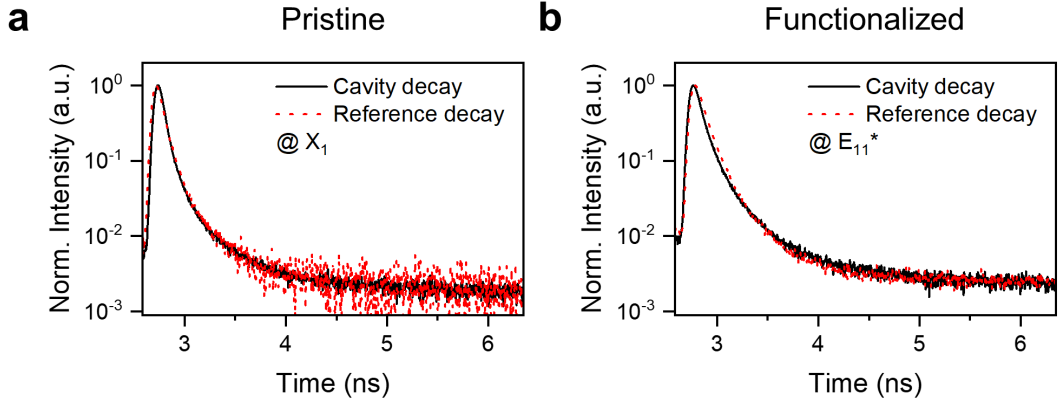

**Figure S11.** Fluorescence decay of pristine (at $X_1$ emission) and functionalized (at $E_{11}^*$ emission) SWCNTs in a cavity and for a reference film.

**Polariton emission at $k_\parallel = 0$ as a function of detuning**

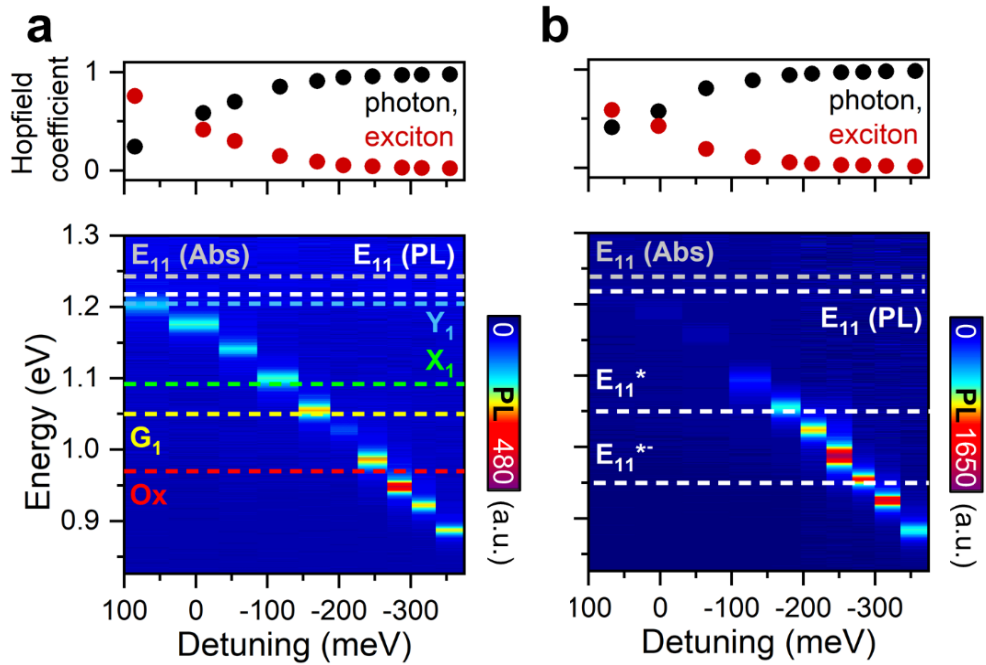

**Figure S12.** Emission intensities of the LP at $k_\parallel = 0$ as a function of detuning for (a) a microcavity filled with pristine SWCNTs, and (b) a microcavity filled with functionalized SWCNTs. The corresponding Hopfield coefficients of the LP for excitons (red dots) and photons (black dots) are given for comparison. The respective photoluminescence sidebands and $sp^3$ defects are indicated as dashed lines for comparison. For cavities tuned around 0.98 eV the λ mode overlaps with the excitation wavelength (640 nm).

**Excitation efficiency correction for population analysis**

To account for the change in excitation efficiency for the different cavities, we simulated the electric field intensity within each structure for the employed excitation wavelength of 640 nm. Since the laser linewidth (±2 nm) is small compared to the cavity resonance (±9 nm), we can neglect its spectral shape. To calculate the electric field, we used the transfer matrix method as described in Ref. 2. Firstly, we matched the mode positions of the experimental reflectivity data by adjusting the SWCNT thickness and holding the metal mirror and oxide spacer thicknesses fixed. With the obtained thicknesses we simulated the electric field and calculated the overlap with the SWCNT layer (**Figures S13**). The resulting overlaps were normalized to the smallest field overlap (1.83) to obtain the correction factors (**Figure S14**). The correction factors were multiplied with the populations of the different samples according to their spacer thickness.

We estimated the uncertainties added by correcting the population data (**Figure 6**) as follows. The highest uncertainty of the simulation input is the SWCNT layer thickness, which is given by the standard deviation of the layer thicknesses used to match the experimental data, that is, 1.8 nm. This uncertainty leads to an error in the LP position of the population data, because the correction factors are mapped using the spacer thickness ($th_{Spacer}$). Additionally, close to the λ mode, even a small change in SWCNT layer thickness affects the excitation efficiency and hence the population strongly. To account for both, we weighted the standard deviation of the SWCNT layer thickness ($th_{SWNCT}$) with the slope of a Bézier interpolation of the correction factors (**Figure S14**).

$$\Delta(th_{SWCNT}) = STD(th_{SWCNT}) \cdot \left(1 + 100 \cdot \left|\frac{\partial \text{BézierInterpol.}}{\partial th_{Spacer}}\right|\right)$$

We converted this uncertainty to the uncertainty in LP position as follows: Firstly, we calculated the optical cavity length ($th_{cav}$) in TM polarization for the strongly coupled $\lambda/2$ mode (the refractive indices $n$ can be assumed constant for the considered spectral range).

$$th_{cav} = (th_{SWCNT} \cdot 2 \cdot n_{SWCNT,TM}) + (th_{spacer} \cdot 4 \cdot n_{Spacer,TM})$$

Secondly, we calculated the respective error, that results from the uncertainty in SWCNT thickness

$$\Delta th_{cav} = (\Delta th_{SWCNT} \cdot 2 \cdot n_{SWCNT,TM})$$

Using error propagation, we converted the uncertainty of the cavity thickness into energy.

$$\Delta th_{cav}[eV] = (hc/th_{cav}^2) \cdot \Delta th_{cav}$$

We assume $\Delta th_{cav}$ to be a good measure for the error of the LP position given in **Figure 6**.

Additionally, the correction factor itself should have an uncertainty due to the previously made approximations. The error should scale with the magnitude of the correction factor. However, this error can only be estimated heuristically. We chose it to be the difference of each correction factor to a B-spline interpolation of the correction factors (**Figure S14**). Together with the standard deviation for averaging over $\pm 1.5°$ around $k_\parallel = 0$, this yielded the error in population shown in **Figure 6**.

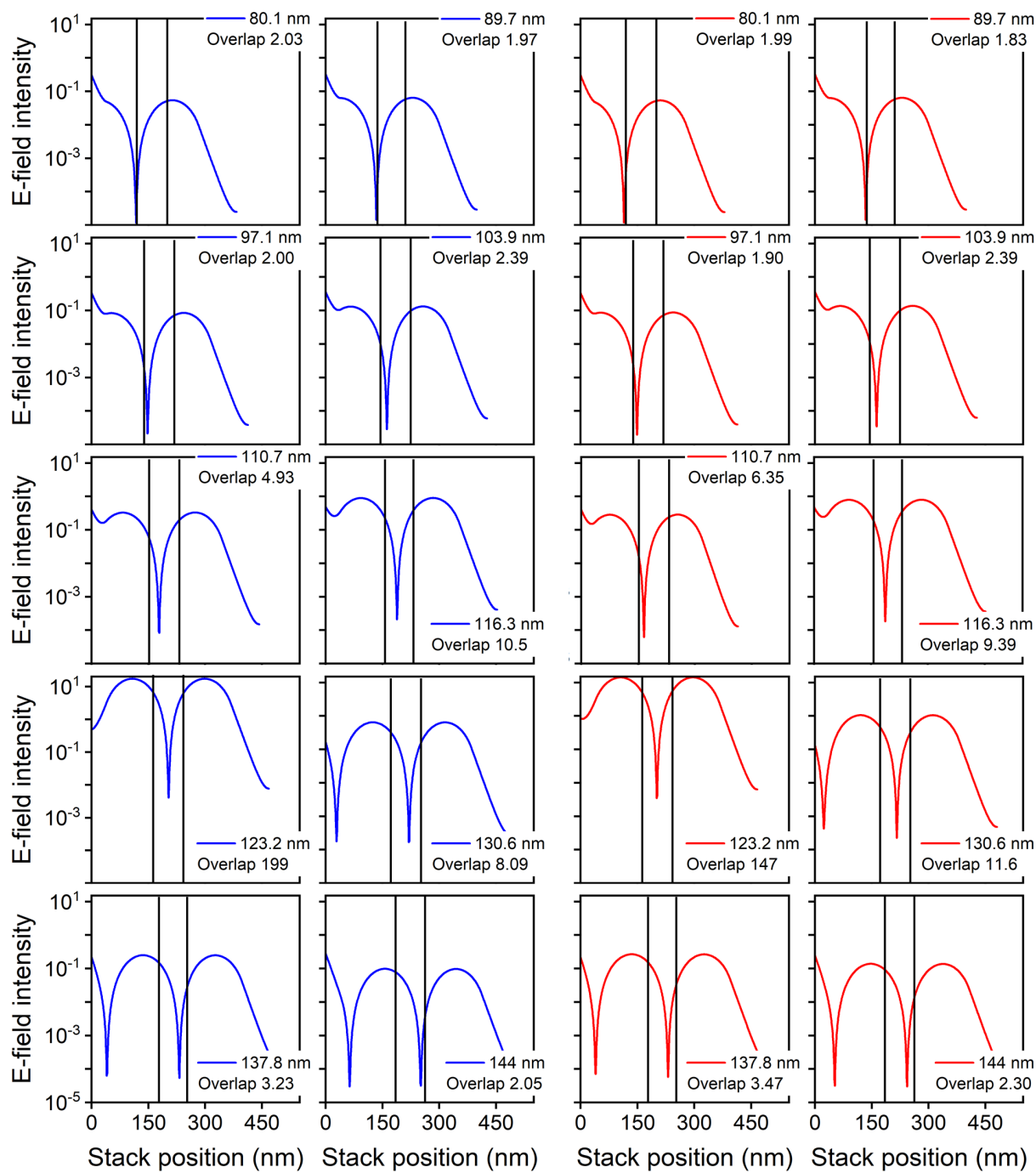

**Figure S13**. Transfer matrix simulation of the electric field intensity as a function of stack position for cavities with pristine (blue) and functionalized SWCNTs (red). The position of the SWCNT layer is indicated by black solid lines. The overlap was calculated as the integral of the electric field intensity within the SWCNT layer.

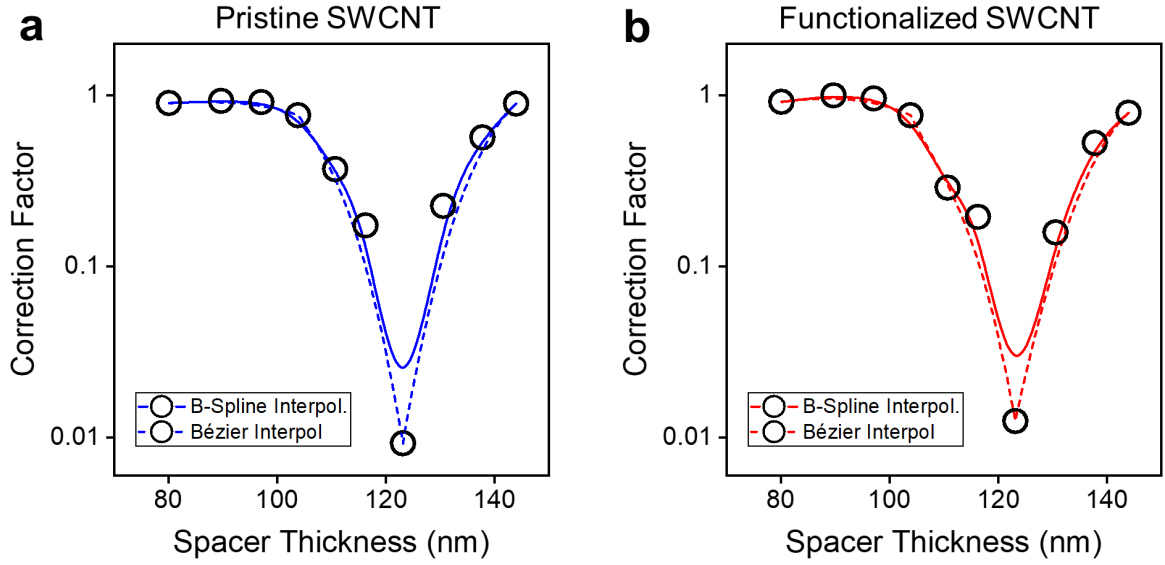

**Figure S14**. Excitation efficiency correction factors for cavities with pristine (a) and functionalized SWCNTs (b). To estimate the error of the correction, the factors were interpolated with a B-spline (solid lines) and a Bézier curve (dashed lines). The highest population reduction is obtained for the cavity detuning at which the λ mode is in resonance with the laser excitation wavelength.

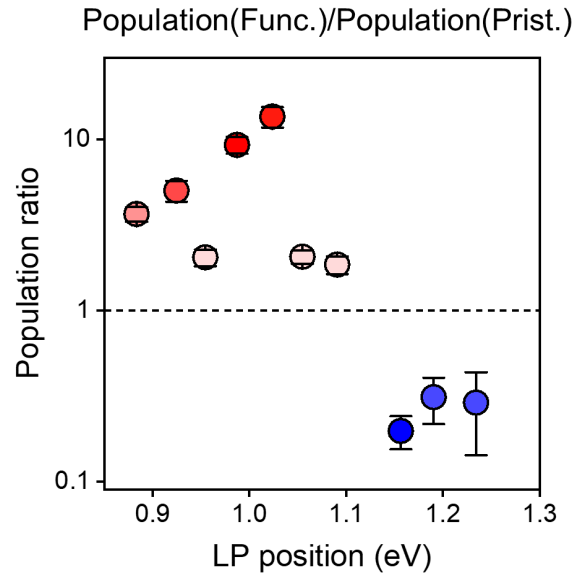

**Figure S15**. Population ratio of the LP population for cavities with functionalized SWCNTs divided by the LP population for cavities with pristine SWCNTs as a function of LP position. The color code is given as in **Figure 6** for comparison.

# LP population as a function of defect emission

To test how the LP population changes with the number of occupied $sp^3$-defect states, we recorded the power dependence of the PL emission of a cavity tuned to the $E_{11}^*$ transition and the PL emission of a functionalized (6,5) SWCNT reference film. Note that the $E_{11}^*$ emission saturates earlier with pump power than the $E_{11}$ emission for functionalized SWCNTs as previously described by Berger *et al.*[1]

To excite the SWCNT layer of both samples with the same power, we accounted for the power loss at the top mirror by measuring the transmission of the excitation source (640 nm laser diode, Coherent OBIS, continuous wave) through a 25 nm gold reference film. The result was extrapolated to the 40 nm top mirror thickness using Beer's law. With this correction we estimated the power loss to be 95 % at 640 nm for the top mirror. We collected the angle-dependent polariton PL and the PL of the reference film at the corresponding excitation powers, the data is shown in **Figure S16a** and **S16b**. After fitting the reflectivity spectrum of the polaritons using the coupled oscillator model we corrected the LP emission at $k_\parallel = 0$ for the LP fraction to extract the polariton population. The $E_{11}^*$ emission intensity was obtained by taking the peak area (**Figure S16b**, shaded area) and subtracting the noise level. For the investigated excitation powers, the LP population at $k_\parallel = 0$ shows a nearly linear dependence on the $E_{11}^*$ emission intensity. Hence, we attribute the change in polariton population directly to the change in $E_{11}^*$ emission, which further confirms the radiative pumping mechanism.

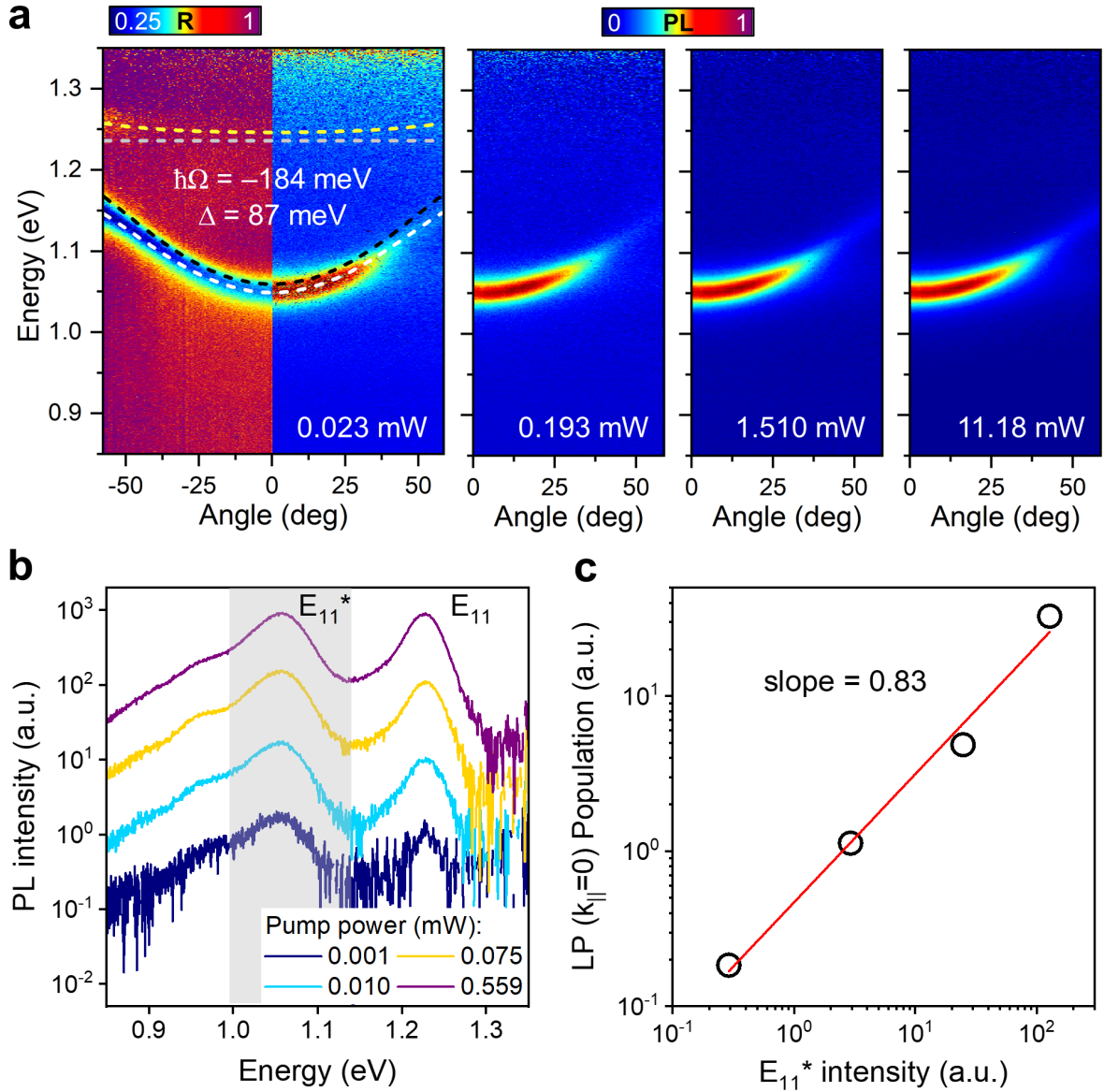

**Figure S16.** (a) Angle-dependent reflectivity and normalized PL spectra for a metal-clad cavity with functionalized SWCNTs tuned to the $E_{11}^*$ transition. The PL spectra are shown for four different excitation powers. (b) PL spectra of a functionalized SWCNT reference film for four equivalent excitation powers with regard to the cavity PL given in (a) (corresponding to 95 % power loss at the top mirror). (c) LP population at $k_\parallel = 0$ given in (a) as a function of the $E_{11}^*$ emission intensity given in (b) for four equivalent excitation powers. The film thickness of the functionalized SWCNT layer was about 80 nm for the cavity and reference.